\documentclass{aa}
\usepackage{graphics,epsfig,amsmath,amssymb,amstext}

\def\d {\mathrm{d}}
\def\Ep {\hbox{$E^{\prime}$}}

\begin{document}

    \thesaurus{}

    \title{Photon Yields of Energetic Particles in the Interstellar
    Medium: an Easy Way to Calculate Gamma-ray Line Emission}

    \author{E. Parizot\inst{1} \and R. Lehoucq\inst{2}}

    \offprints{E.~Parizot}

    \institute{Institut de Physique Nucl\'eaire d'Orsay,
    IN2P3-CNRS/Universit\'e Paris-Sud, 91406 Orsay Cedex, France \and
    DAPNIA/Service d'astrophysique, CEA-Saclay, 91191 Gif-sur-Yvette,
    France }

    \date{Received 9 July 2001; accepted 11 December 2001}

    \authorrunning{E. Parizot and R. Lehoucq}
    \titlerunning{EP Photon Yields in the ISM}

    \maketitle

    \begin{abstract}
	
	A $\gamma$-ray line production calculation in astrophysics
	depends on i) the composition and energy source spectrum of
	the energetic particles, ii) the propagation model, and iii)
	the nuclear cross sections.  The main difficulty for model
	predictions and data interpretation comes from the fact that
	the spectrum of the particles which actually interact in the
	ISM -- the \textit{propagated spectrum}, is not the same as
	the \textit{source spectrum} coming out of the acceleration
	site, due to energy-dependent energy losses and nuclear
	destruction.  We present a different approach to calculate
	$\gamma$-ray line emission, based on the computation of the
	total number of photons produced by individual energetic
	nuclei injected in the interstellar medium at a given energy. 
	These photon yields take into account all the propagation
	effects once and for all, and allow one to calculate quickly
	the $\gamma$-ray line emission induced by energetic particles
	in any astrophysical situation by using directly their source
	spectrum.  Indeed, the same photon yields can be used for any
	source spectrum and composition, as well as any target
	composition.  In addition, these photon yields provide visual,
	intuitive tools for $\gamma$-ray line phenomenology.
    
	\keywords{Cosmic rays; Gamma rays: theory; Nuclear reactions}

    \end{abstract}

    \section{Introduction}
    
    High energy astrophysics is experiencing a considerable
    development, notably through the operation of a new generation of
    gamma-ray observatories, both on the ground and onboard
    satellites.  The analysis of gamma-ray emission from compact and
    diffuse sources is one of the most valuable ways to study high
    energy processes in the universe, and it is expected to put
    stronger and stronger constraints on the models in the near
    future.  Among the processes of interest, the production of
    gamma-ray lines by energetic particles (EPs) interacting in the
    interstellar medium (ISM) has received increased attention in the
    last few years, in connection with the data of the Compton
    Gamma-Ray Observatory as well as the forthcoming satellite
    INTEGRAL. In addition to the study of the high energy sources, the
    study of EP interactions is important for the understanding of the
    EPs themselves, of which the Galactic cosmic rays (GCRs) are one
    of the main components.

    Information about the acceleration sites and processes is also
    provided by the determination of the EP energy spectrum and
    chemical composition, which can be derived, in principle, from the
    measurement of gamma-ray line ratios and profiles.  However, the
    information contained in the gamma-ray observational data relates
    to the spectrum and composition of the \textit{propagated}
    particles (i.e. those who actually interact in the ISM), not the
    \textit{source} particles.  The difference arises from the fact
    that the EPs experience various types of interactions while they
    `propagate' from their source to the place where they produce
    gamma-ray lines.  In particular, they lose energy through
    Coulombian interactions in a way which depends on both their
    energy and chemical nature, so that the propagated population of
    EPs is not identical to the source population, freshly coming out
    of the acceleration process.
    
    Ideally, one would divide the whole process into three successive
    stages (e.g. Parizot and Lehoucq, 1999): particle acceleration,
    propagation and interaction, with the gamma-ray production arising
    during the last stage only.  In reality, of course, the reactions
    leading to gamma-ray production occur all the time, from the
    injection of the EPs into the ISM until they have slowed down to
    energies below the interaction thresholds.  It is therefore
    necessary to sum the contributions of all the instants following
    acceleration (the gamma-ray emission occurring during acceleration
    itself can usually be neglected, except for the rarest, highest
    energy particles, which spent a long time in the accelerator).  In
    steady state situations, this is equivalent to calculating the
    equilibrium distribution of EPs and integrating the relevant cross
    sections over this so-called propagated distribution.
    
    From a technical point of view, the most difficult part consists
    in calculating the propagated spectrum, taking into account the
    energy losses and the energy-dependent escape time of the
    particles out of the confinement region, where the gamma-ray
    production is evaluated.  This is what prevents a straightforward
    calculation of the expected gamma-ray line fluxes from the
    knowledge of the nuclear cross sections and the EP source
    distribution.  Therefore, we propose here to work out this step
    once and for all, in the case of a steady state and a thick
    target, by calculating the integrated effect of energy losses on
    individual particles injected at any energy in the ISM. In
    Sect.~\ref{sec.universality}, we shall justify the fact that the
    `propagation step' in a standard $\gamma$-ray line emission
    calculation can be `factorized out' and calculated separately,
    independently of the EP source spectrum and composition.  This is
    true if the metallicity of the propagation medium does not exceed
    several tens of times the solar metallicity, as in most of the
    astrophysically sensible situations.  As a result, we shall obtain
    the absolute $\gamma$-ray yields of energetic nuclei as a function
    of their initial energy, from which the $\gamma$-ray line emission
    induced by EPs of any spectrum and composition can be
    straightforwardly calculated.  In addition to making $\gamma$-ray
    line calculations much easier, these absolute yields (or particle
    efficiencies for $\gamma$-ray line production) considerably help
    phenomenological interpretation of the observational data, as
    these yields only need to be convolved with the EP \textit{source}
    spectra, rather than \textit{propagated} ones.

    \section{Standard gamma-ray line calculations}\label{sec.Standard}
    
    \begin{figure*}
       \centering
       \includegraphics[width=8cm]{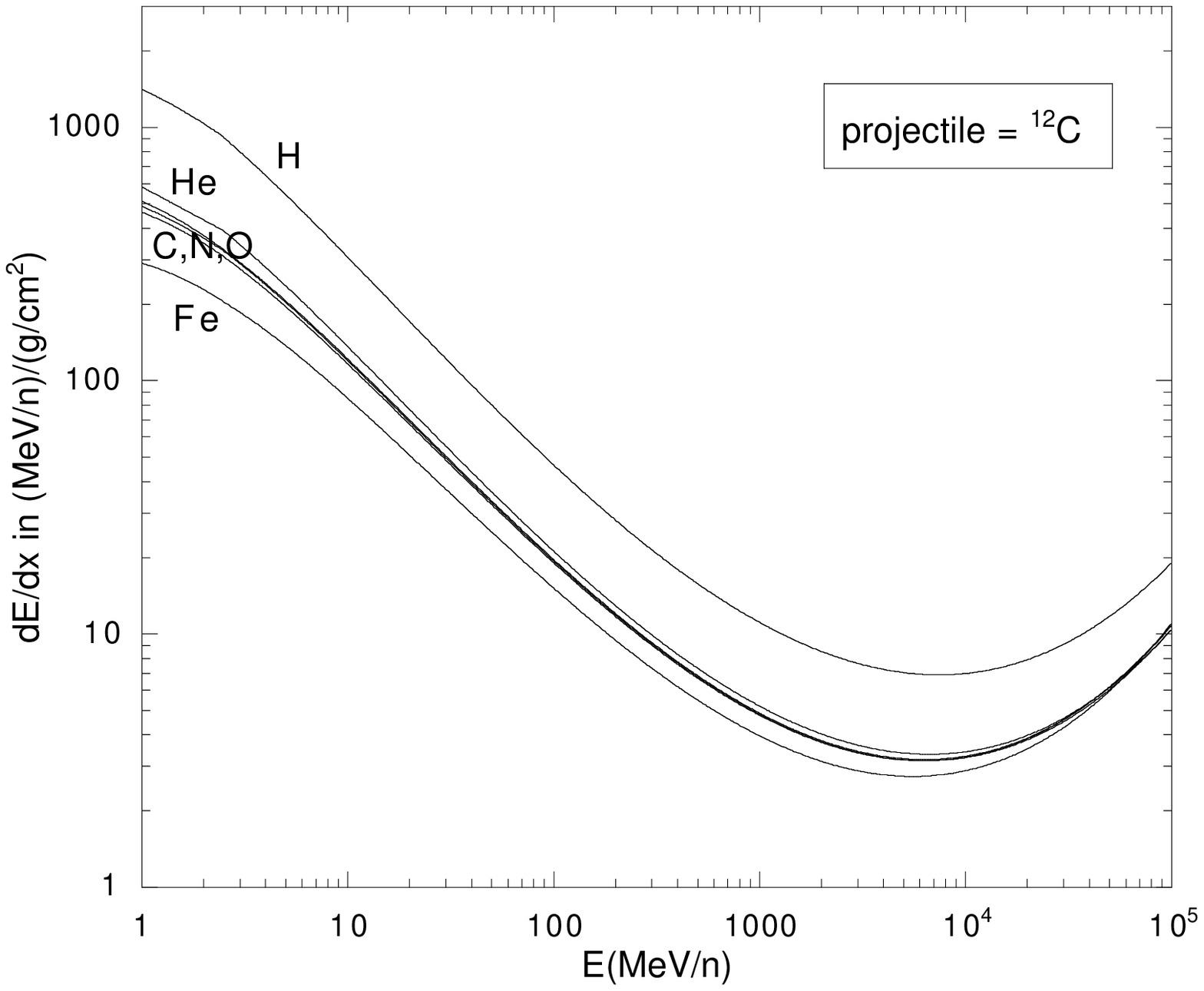}\hspace{0.2cm}
       \includegraphics[width=8cm]{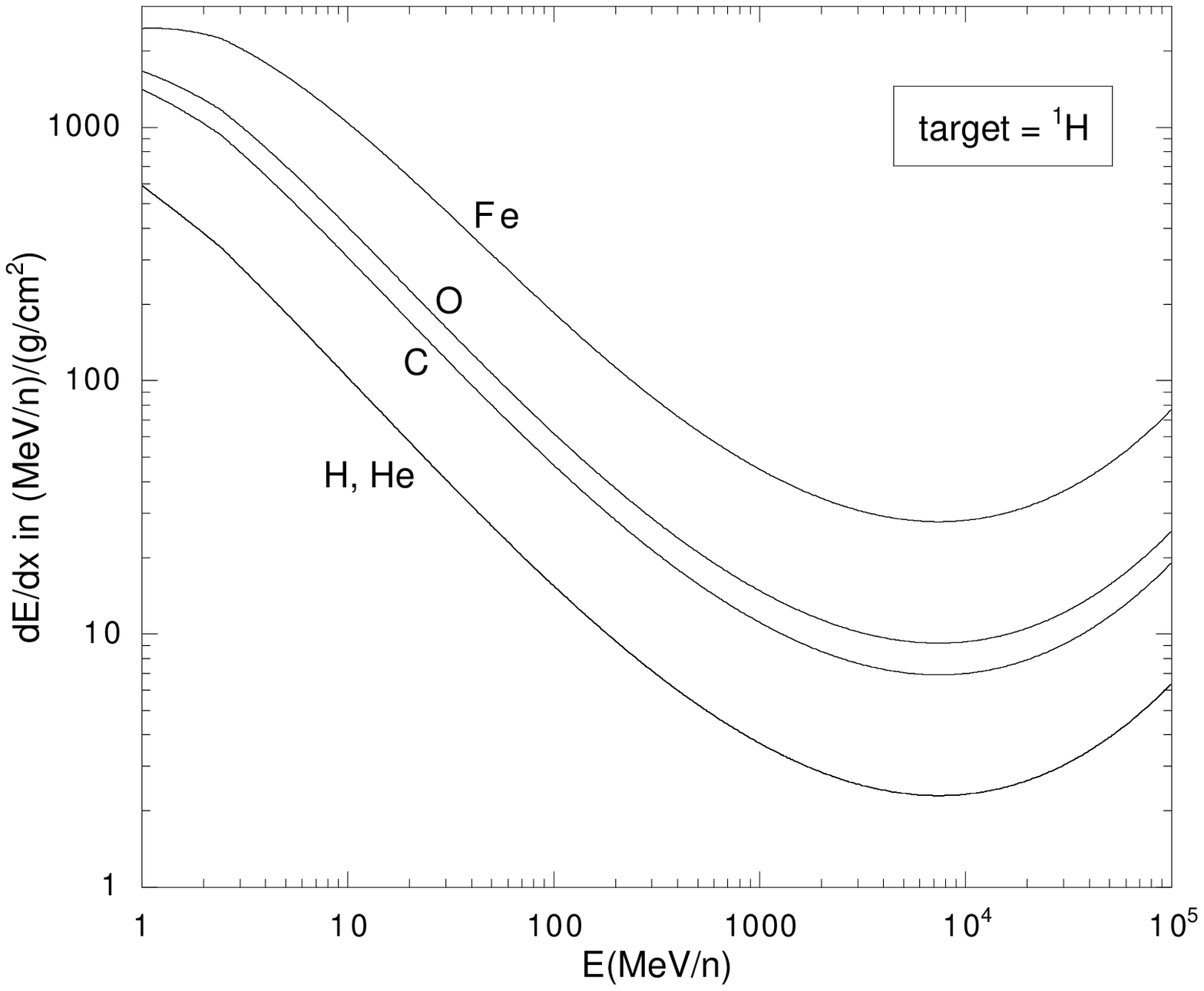}
       \caption{Energy losses per unit grammage of matter passed
       through, in $(\mathrm{MeV/n})/(\mathrm{g}\,\mathrm{cm}^{-2})$:
       on the left, for a $^{12}$C nucleus in various chemically pure
       propagation media indicated by the labels; on the right, for
       various nuclei indicated by the labels in a propagation medium
       made of pure Hydrogen.}
       \label{energyLosses}
    \end{figure*}
    
    In order to calculate the gamma-ray line emission in a given
    region of the ISM, one needs to integrate the nuclear excitation
    cross sections over the local flux of energetic particles, and sum
    all the contributions to each gamma-ray line.  If $i$ represents
    the projectile, $j$ the target nucleus, and $k$ the excited
    nucleus produced in the interaction between $i$ and $j$, or
    equivalently the `photon species' emitted through nuclear
    de-excitation, the $\gamma$-ray line emission rate, in
    $\mathrm{ph}/\mathrm{cm}^{3}/\mathrm{s}$, from all the $i + j
    \rightarrow k$ nuclear reactions, reads:
    
    \begin{equation}
	\begin{split}
	\frac{\d N_{k}}{\d t} &= \sum_{i,j}\int_{0}^{+\infty}
	N_{i}(E) \left[n_{j} \sigma_{i,j;k}(E)v(E)\right]\d E \\ &=
	\sum_{i,j}\int_{0}^{+\infty}\Phi_{i}(E)n_{j}
	\sigma_{i,j;k}(E)\d E,
	\end{split}
	\label{eq:TauxProdStat}
    \end{equation}
    where $N_{i}(E)$ is the spectral density of the projectiles $i$,
    in $\mathrm{cm^{-3}}(\mathrm{MeV/n})^{-1}$, $\Phi_{i}(E) =
    N_{i}(E)\times v(E)$ is the corresponding flux, in
    $\mathrm{cm}^{-2}\mathrm{s}^{-1}(\mathrm{MeV/n})^{-1}$, $v(E)$ is
    the velocity of the projectiles (independent of the nuclear
    species, $i$, if the energy $E$ is expressed in MeV/n), $n_{j}$ is
    the number density of the target nuclear species $j$, and
    $\sigma_{i,j;k}$ is the cross section for the reaction $i + j
    \rightarrow k$.  For instance, $k$ might represent photons from
    the $^{12}\mathrm{C}$ de-excitation line at 4.44~MeV, produced by
    the reaction $p + \mathrm{^{12}C} \longrightarrow
    \mathrm{^{12}C}^*$ or $\alpha + \mathrm{^{16}O} \longrightarrow
    \mathrm{^{12}C^*}$.

    Assuming that the nuclear excitation cross-sections are known, the
    main challenge is to estimate the EP fluxes, for each nuclear
    species, in the gamma-ray source.  This depends on the
    acceleration process at the origin of the injection of EPs in the
    ISM, and on what happens to the particles once they leave the
    accelerator, which involves the energy losses, the rate of escape
    from the region considered, and the particle destruction in
    inelastic nuclear processes.  As far as the acceleration is
    concerned, it can be characterized here by the so-called
    \textit{injection function}, $Q_{i}(E)$, which gives the number of
    particles of species $i$ injected at energy $E$ in the ISM (i.e.
    leaving the acceleration process and not being further accelerated
    afterwards), in
    $\mathrm{cm^{-3}}\mathrm{s}^{-1}(\mathrm{MeV/n})^{-1}$.  The
    function $Q_{i}(E)$ will either be taken as the outcome of some
    particular acceleration model (e.g. diffusive shock acceleration),
    or phenomenologically postulated so as to reproduce some
    particular observation (e.g. from INTEGRAL data).
    
    Most studies so far have assumed that the shape of the injection
    spectrum, $Q_{i}(E)$, was independent of the nuclear species, and
    could be re-written as $\alpha_{i}\bar{Q}(E)$, where the isotopic
    abundances $\alpha_{i}$ characterize the chemical composition of
    the EPs \textit{at the source}.  However, this simplification is
    not required and one will allow here for a different spectrum for
    each nuclear species, which is equivalent to an energy dependent
    EP composition.
    
    In order to use Eq.~(\ref{eq:TauxProdStat}) to calculate the
    gamma-ray emission produced in the region under consideration, one
    needs to derive the EP fluxes, $\Phi_{i}(E)$ or $N_{i}(E)$, from
    the injection functions, $Q_{i}(E)$, supposed known.  The standard
    way to do this has been described in Parizot and Lehoucq (1999,
    and references therein) for the general case where the injection
    function as well as the conditions of propagation are
    time-dependent.  It consists in solving the so-called propagation
    equation, which takes the following form in the stationary case:
    
    \begin{equation}
	\frac{\partial} {\partial E}(\dot E_{i}(E) N_{i}(E) ) =
	Q_{i}(E) - \frac{N_{i}(E)}{\tau_{i}^{\mathrm{tot}}(E)}.
	\label{eq:StatPropEqua}
    \end{equation}

    Here, $\dot E_{i}(E)$ is the energy loss function, in
    $(\mathrm{MeV/n})\,\mathrm{s}^{-1}$, giving the rate of energy
    loss for nuclei of species $i$ in the considered propagation
    medium, and $\tau_{i}^{\mathrm{tot}}(E)$ is the total `loss time'
    taking into account nuclear destruction and particle escape out of
    the region considered.  It can be expressed in terms of the total
    inelastic cross sections for nuclei of species $i$ in a medium
    made of $j$ nuclei alone, $\sigma_{i,j}^{\mathrm{dest}}(E)$, and
    the mean escape time, $\tau_{i}^{\mathrm{esc}}(E)$, as:
    
    \begin{equation}
	\frac{1}{\tau_{i}^{\mathrm{tot}}(E)} =
	\frac{1}{\tau_{i}^{\mathrm{esc}}(E)} +
	\sum_{j}n_{j}\sigma_{i}^{\mathrm{dest}}(E)v(E),
	\label{eq:totalLossTime}
    \end{equation}
    where $n_{j}$ is as above the number density of nuclei of species
    $j$ in the propagation medium.
    
    In Eq.~(\ref{eq:StatPropEqua}), the injection function $Q_{i}(E)$
    acts as a source term, while the equilibrium EP distribution
    $N_{i}(E)$ is what we want to calculate.  The formal solution of
    the stationary propagation equation reads:
    
    \begin{equation}
	N_{i}(E) = \frac{1}{|\dot{E}_{i}(E)|}\int_{E}^{+ \infty}
	Q_{i}(E_{\mathrm{in}})\mathcal{P}_{i}(E_{\mathrm{in}},E)
	\d E_{\mathrm{in}},
	\label{eq:StatSolution}
    \end{equation}
    where $\mathcal{P}_{i}(E_{\mathrm{in}},E)$ can be interpreted as
    the survival probability, in the propagation medium considered, of
    a particle injected at energy $E_{\mathrm{in}}$ and losing energy
    down to energy $E$.  It obviously depends on the total loss time
    at each energy between $E_{\mathrm{in}}$ and $E$ and the energy
    loss function, and can be expressed as follows (see Parizot and
    Lehoucq, 1999):
    
    \begin{equation}
	\mathcal{P}_{i}(E_{\mathrm{in}},E) = \exp\Bigg(-
	\int_{E_{\mathrm{in}}}^{E}\frac{\d\Ep}{\dot{E}_{i}(\Ep)
	\tau_{\mathrm{tot},i}(\Ep)}\Bigg).
	\label{eq:survivalProba}
    \end{equation}

    \section{Approached EP propagation universality}
    \label{sec.universality}
    
    \begin{figure*}[t]
       \centering
       \includegraphics[width=8cm]{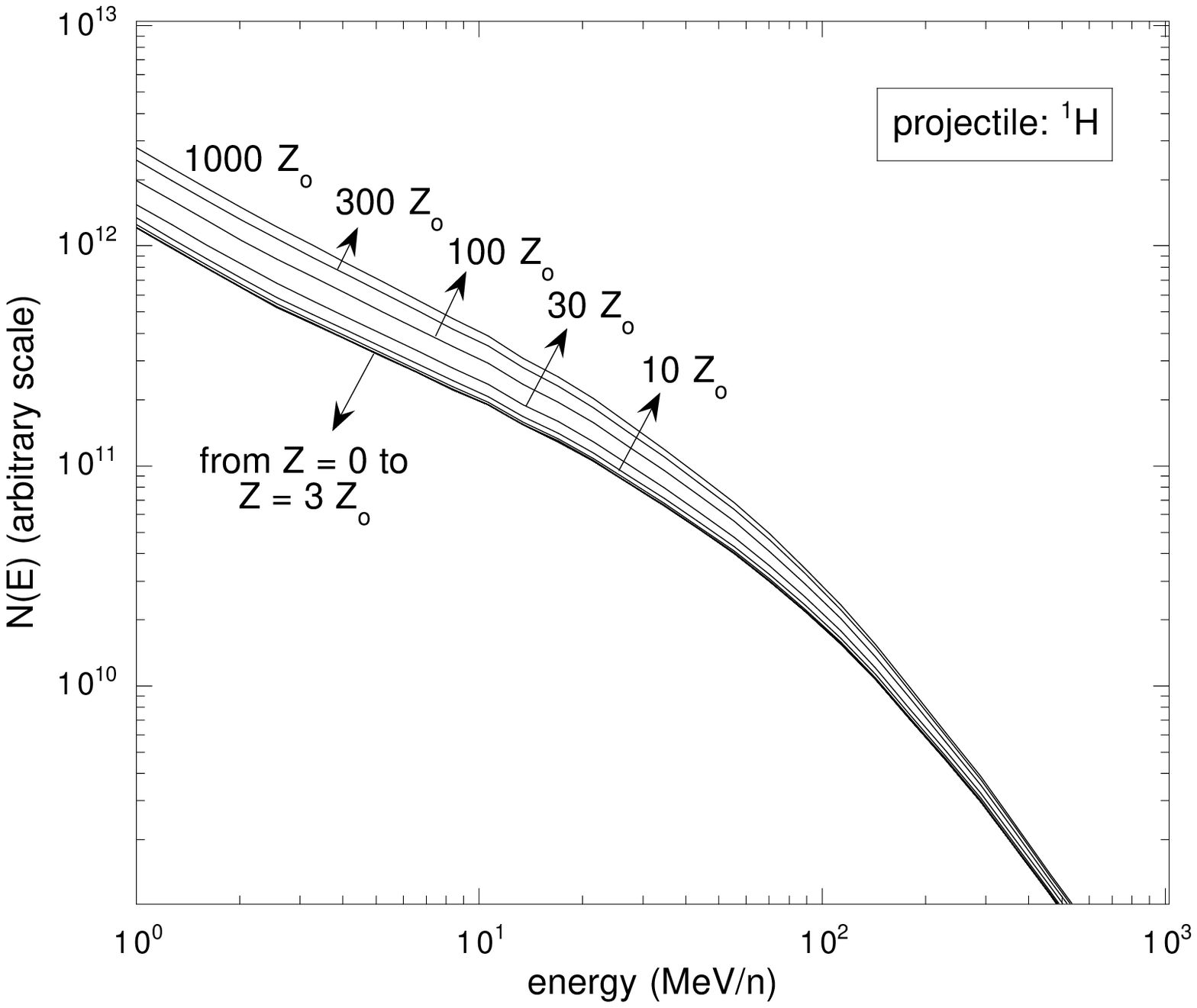}\hspace{0.2cm}
       \includegraphics[width=8cm]{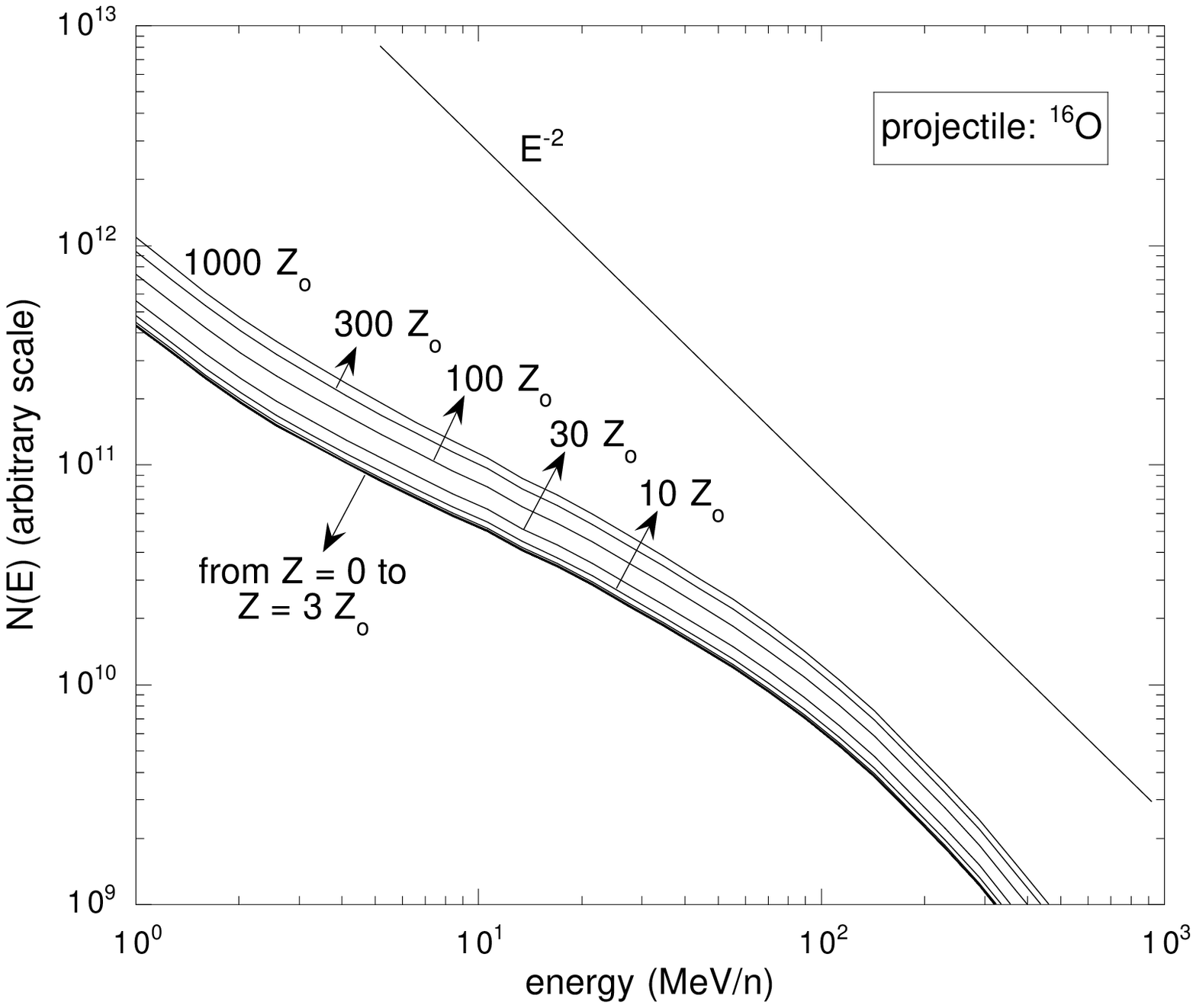}
       \caption{Stationary energy spectra of $^{1}$H (left) and
       $^{16}$O (right) nuclei injected with a simple power-law
       spectrum in energy of index 2, after propagation in different
       media.  The spectra corresponding to propagation media with a
       metallicity up to 10~$\mathrm{Z}_{0}$ are virtually
       indistinguishable from the propagated spectrum in a metal-free
       gas.}
       \label{propagatedSpectra}
    \end{figure*}

    Equation~(\ref{eq:TauxProdStat}) allows one to calculate the
    photon emission rate for any gamma-ray de-excitation line for
    which the production cross sections are known.  It is based on the
    computation of the EP distribution function given by
    Eq.~(\ref{eq:StatSolution}), which in turn requires the knowledge
    of the energy losses and the escape and destruction times.
    
    \subsection{Energy loss rates}
    
    Let us first consider the energy loss rate, $\dot{E}_{i}(E)$, for
    particles of species $i$.  It can be expressed as a sum over the
    nuclear species present in the propagation medium:

    \begin{equation}
	\dot{E}_{i}(E) =
	\sum_{j}n_{j}\times\frac{\d E}{\d x}\bigg|_{j}\times v(E),
	\label{eq:energyLoss}
    \end{equation}
    where $(\d E/\d x)_{j}$ is the energy loss per unit grammage in a
    medium of pure $j$ nuclei, in
    $(\mathrm{MeV/n})/(\mathrm{g}\,\mathrm{cm}^{-2})$.  In this
    expression, $(\d E/\d x)_{j}$ is a `universal function' of energy
    which can ideally be calculated from first physical principles or,
    failing that, extrapolated from laboratory measurements, while the
    astrophysics comes in the chemical composition, the degree of
    ionization and the density of the propagation medium, $n_{j}$. 
    Some examples of energy loss functions used in this paper are
    shown in Fig.~\ref{energyLosses}.  They have been calculated using
    the program of J. Kiener (1994), based on the modified Bethe
    formula taking into account the effective charge of the
    projectile.  This program implements the Ziegler tables for the
    various stopping powers, as corrected by Hubert et al.  (1989)
    according to a semi-empirical procedure, where a new
    parameterization for the effective charge is deduced from a very
    large set of experimental stopping power values in the range
    3--80~MeV/n.  The typical error in the energy loss function is
    between 2\% and 10\%, i.e. less than the uncertainty about the
    nuclear excitation cross-sections.
    
    The energy range of interest for the calculation of gamma-ray line
    emission is between a few MeV/n (corresponding to the nuclear
    excitation thresholds) and a few hundreds of MeV/n.  At higher 
    energy, the contribution of the EPs to the gamma-ray line emission
    is small, because of their reduced number (decreasing power-law
    source spectrum) as well as because they are destroyed by nuclear
    reactions before they reach the peaks of the nuclear excitation
    cross sections, as further discussed below.

    In all the calculations presented here, the ambient medium is
    assumed neutral, which may not be appropriate for a number of
    astrophysical situations.  However, one can estimate that the
    effect of an ambient ionized medium is small, except for the
    lowest energies.  The reason why the energy losses depend on the
    ionization state of the propagation medium is that it is more
    difficult for an energetic ion to capture a free electron than to
    capture an electron from an atom at rest.  Indeed, in the latter
    case, the orbital motion of the electron reduces the velocity
    difference between the energetic ion and the electron, and thereby
    facilitates capture.  As a consequence, the equilibrium between
    electron stripping and electron capture depends on the ambient
    medium, and an energetic atom is on average more ionized when it
    travels through a plasma than through a neutral medium (Chabot et
    al., 1995a).  This results in a higher effective charge, and thus
    a higher stopping power (or larger energy losses).  However, when
    the projectile is too energetic, its relative velocity with even
    orbital electrons is too high for charge exchange to be efficient
    anyway.  Therefore, the difference between an ionized and a
    neutral ambient medium becomes negligible, and the energy losses
    are almost identical.  From the quantitative point of view, the
    stopping power of a plasma is higher than that of a neutral medium
    by a factor of about 40 below 100~keV/n, but only 2 or 3 at a few
    MeV/n, and their difference is negligible above 100~MeV/n (Hoffman
    et al., 1994; Chabot et al., 1995b).  Therefore, we will assume
    throughout that the propagation medium is neutral.
    
    As far as the chemical composition is concerned, in most
    astrophysically relevant cases the propagation medium will be but
    the ISM, whose composition is relatively well known (Anders and
    Grevesse, 1989).  Now, although the heavy elements are of course
    crucial to the calculation of gamma-ray de-excitation line
    emission, the ISM is so much dominated by H and He nuclei that one
    can neglect all other elements in Eq.~(\ref{eq:energyLoss}), and
    calculate the energy loss rate as if the ISM were made simply of
    91\% of H and 9\% of He (by number).  To demonstrate this, we have
    calculated the propagated (equilibrium) spectrum of the different
    energetic nuclei subject to energy losses in media of various
    metallicity.  Results are shown in Fig.~\ref{propagatedSpectra}
    for metallicities ranging from 0 (H and He only) to 1000 times
    solar.  A significant change in the particle propagated spectrum
    can only be noticed for ambient metallicities larger than ten
    times the solar metallicity, $Z_{0}$.  Since most of the
    astrophysically relevant media are not that rich in metals, we
    will assume that the propagation of the EPs is independent of
    metallicity.  Note that even pure SN ejecta have a metallicity
    less than 30 times $Z_{0}$, so that even in such a metal-rich
    medium, neglecting the interaction of the EPs with the metals as
    they propagate through the ambient medium will lead to an error
    smaller than 20\% on the propagated particle distribution,
    $N_{i}(E)$, and thus also on the gamma-ray line production rates.
    
    \subsection{Total inelastic cross sections and survival
    probabilities}
    
    The second crucial physical ingredient necessary to calculate the
    EP propagated spectra is the total `catastrophic loss time',
    $\tau_{i}^{\mathrm{tot}}(E)$, including both nuclear destruction
    and escape (see Eq.~(\ref{eq:totalLossTime})).  The latter will be
    neglected here, which amounts to say that we consider a
    distribution of EPs interacting with a thick target.  This is well
    justified for the relatively low energy particles which are
    responsible for most of the gamma-ray line emission, as their
    range is of the order of $1\,\mathrm{g/cm}^{2}$, i.e.
    significantly less than the typical escape grammage of cosmic
    rays. One can thus replace the total catastrophic loss 
    timescale, $\tau_{i}^{\mathrm{tot}}(E)$, by the destruction 
    timescale, $\tau_{i}^{\mathrm{D}}(E)$.
    
    Concerning the EP destruction through nuclear reactions, we use
    semi-empirical total inelastic cross sections from Silberberg et
    Tsao (1990).  As for the energy loss function, we have just seen
    that the EP destruction time can be calculated as if the
    propagation medium were made of pure H and He, to an excellent
    approximation (even for relatively high ambient metallicities). 
    In all of our calculations, the main uncertainty comes from the
    relatively poor knowledge of the nuclear cross sections, rather
    than from the use of a `universal' (metal-free) propagation
    medium.
    
   \begin{figure}[t]
       \centering
       \includegraphics[width=8cm]{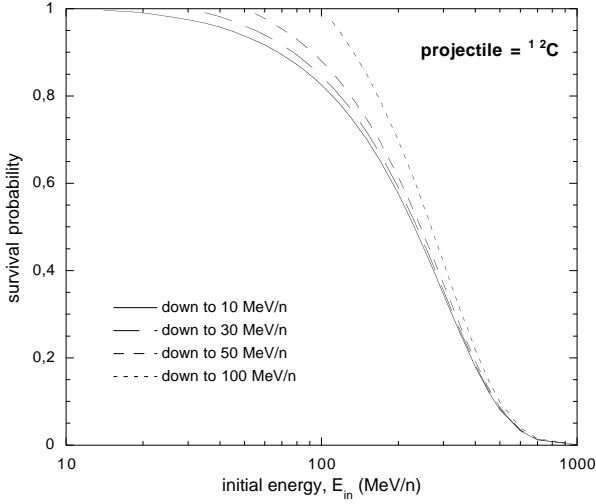}
       \caption{EP survival probability down to 10, 30, 50 and
       100~MeV/n, as a function of the injection energy,
       $E_{\mathrm{in}}$.}
       \label{fig:survivalProba}
   \end{figure}
   
    The joint knowledge of $\dot{E}_{i}(E)$ and
    $\tau_{i}^{\mathrm{tot}}(E)$ allows one to calculate the survival
    probability of an EP injected at any energy $E_{\mathrm{in}}$
    while it loses energy down to the energy $E$ at which it interacts
    with the ISM to produce a gamma-ray.  It can be useful to define
    the energy loss timescale, $\tau_{i}^{\mathrm{loss}}(E)$ for an EP
    of type $i$ as:
    \begin{equation}
	\tau_{i}^{\mathrm{loss}}(E) = E/|\dot{E}_{i}(E)|
	\label{eq:tauLoss}
    \end{equation}
    
    The survival probability given by Eq.~(\ref{eq:survivalProba}) can 
    then be rewritten as:
    \begin{equation}
	\mathcal{P}_{i}(E_{\mathrm{in}},E) =
	\exp\Bigg(-\int_{E}^{E_{\mathrm{in}}}
	\frac{\tau_{i}^{\mathrm{loss}}(\Ep)}{\tau_{i}^{\mathrm{D}}(\Ep)}
	\frac{\d\Ep}{\Ep}\Bigg).
	\label{eq:survivalProbaBis}
    \end{equation}
    
    Fig.~\ref{fig:survivalProba} shows the results obtained in the
    case of a $^{12}$C nucleus.  As can be seen, particles injected at
    energies greater than 1~GeV/n have very little chance to survive
    long enough to reach energies of a few tens of MeV/n, where the
    nuclear excitation cross sections are maximum.  As a consequence,
    their contribution to the total gamma-ray emission rate will be
    negligible, not mentioning the fact that higher energy particles
    are usually less numerous than lower energy ones, for the most
    common, power-law source spectra.  This effect reminds us that
    nuclear excitation is only one of a number of competing nuclear
    processes affecting an energetic particle, so that the efficiency
    of gamma-ray line production depends on the balance between
    several cross sections.

    \section{Gamma-ray yields of individual EPs}
   
    We now make use of the results of the previous section to derive
    an alternate way to calculate gamma-ray line production from EPs. 
    This idea is the following: since particle propagation is
    `universal' (i.e. independent of the physical conditions of the
    propagation medium, as long as metals can be neglected), there
    should be a way to work it out once and for all.  This way is the
    following: instead of calculating the equilibrium EP distribution
    function (after propagation), and then integrate the nuclear
    cross section over this distribution, one can calculate the total
    photon yield, from injection to rest, of one EP of a given kind
    thrown in the ISM at any given initial energy, independently of
    the other particles (i.e. independently of the global spectrum and
    composition of the EPs), and then integrate the individual photon
    yields over the source distribution function of each kind of EPs
    accelerated.  The important point is that the integration is now
    over the \textit{source} distribution function, which is known
    from acceleration models or postulated in a phenomenological
    study, rather than over the \textit{propagated} distribution
    function, which no longer has to be calculated.
   
\begin{figure}
	\includegraphics[width=8.3cm]{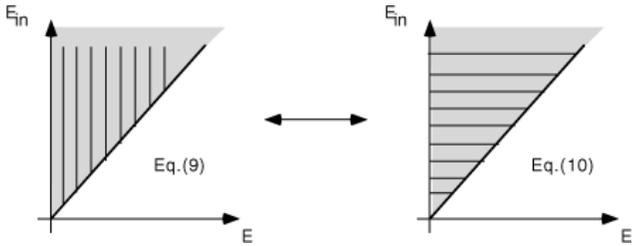}
	\caption{Graphical demonstration of the equivalence between
	Eq.~(\ref{eq:ProdRate1}) and Eq.~(\ref{eq:ProdRate2}): the
	shaded area is the integration domain, divided into vertical
	and horizontal slices, respectively.}
        \label{fig:MathTransform}
    \end{figure}

    From the mathematical point of view, the above idea amounts to a
    simple change of the order of two integrations.  Indeed, combining
    Eqs.~(\ref{eq:TauxProdStat}) and~(\ref{eq:StatSolution}), one can
    rewrite the $\gamma$-ray emission rate as follows (specializing to
    one nuclear reaction for illustration):

    \begin{equation}
        \frac{\d N_{\gamma}}{\d t} =
        \int_{0}^{+\infty}\hspace{-0.5cm}\d E
        \int_{E}^{+\infty}\hspace{-0.5cm}\d E_{\mathrm{in}}
        \frac{n_{0}\sigma(E)v(E)}{|\dot{E}(E)|}Q(E_{\mathrm{in}})
        \mathcal{P}(E_{\mathrm{in}},E),
        \label{eq:ProdRate1}
    \end{equation}
    where $n_{0}$ is the density of the propagation medium and $Q(E)$
    is the EP \textit{source} spectrum.  Now, as is made obvious by
    Fig.~\ref{fig:MathTransform}, this expression can be re-written
    as:

    \begin{equation}
        \frac{\d N_{\gamma}}{\d t} =
        \int_{0}^{+\infty}\hspace{-0.5cm}\d E_{\mathrm{in}}
        \int_{0}^{E_{\mathrm{in}}}\hspace{-0.3cm}\d E\hspace{0.05cm}
        \frac{n_{0}\sigma(E)v(E)}{|\dot{E}(E)|}Q(E_{\mathrm{in}})
        \mathcal{P}(E_{\mathrm{in}},E).
        \label{eq:ProdRate2}
    \end{equation}
   
    Getting the source function, $Q(E_{\mathrm{in}})$, out of the
    integral over $E$, one obtains the following expression for the
    $\gamma$-ray emission rate (adding the contribution of all the
    reactions involved):
    \begin{equation}
	\frac{\d N_{\gamma}}{\d t} = \sum_{i,j}
	\int_{0}^{+\infty}\hspace{-0.2cm}Q_{i}(E_{\mathrm{in}})
	\alpha_{j}\mathcal{N}_{i,j;\gamma}(E_{\mathrm{in}})\d
	E_{\mathrm{in}},
	\label{eq:ProdRateInverted}
    \end{equation}
    where $\alpha_{j} = n_{j}/n_{0}$ is the relative abundance of
    nuclei of species $j$ in the target, and
    \begin{equation}
	\mathcal{N}_{i,j;\gamma}(E_{\mathrm{in}}) =
	\int_{0}^{E_{\mathrm{in}}}
	\frac{n_{0}\sigma_{i,j;\gamma}(E)v(E)}{|\dot{E}(E)|}
	\mathcal{P}_{i}(E_{\mathrm{in}},E) \d E.
	\label{eq:PhotonYield}
    \end{equation}

    \begin{figure*}
	\centering
	\includegraphics[width=8cm]{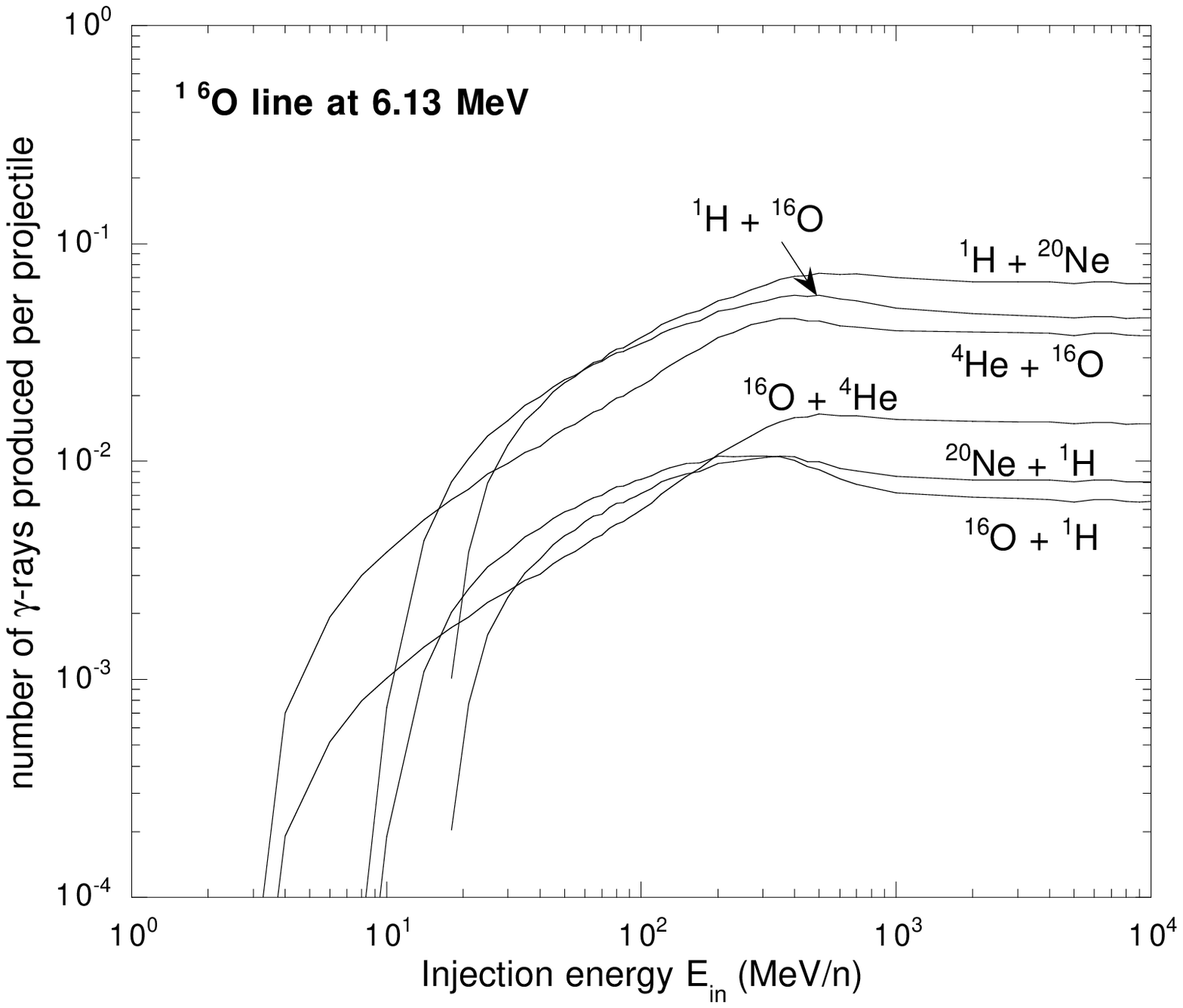}\hspace{0.2cm}
	\includegraphics[width=8cm]{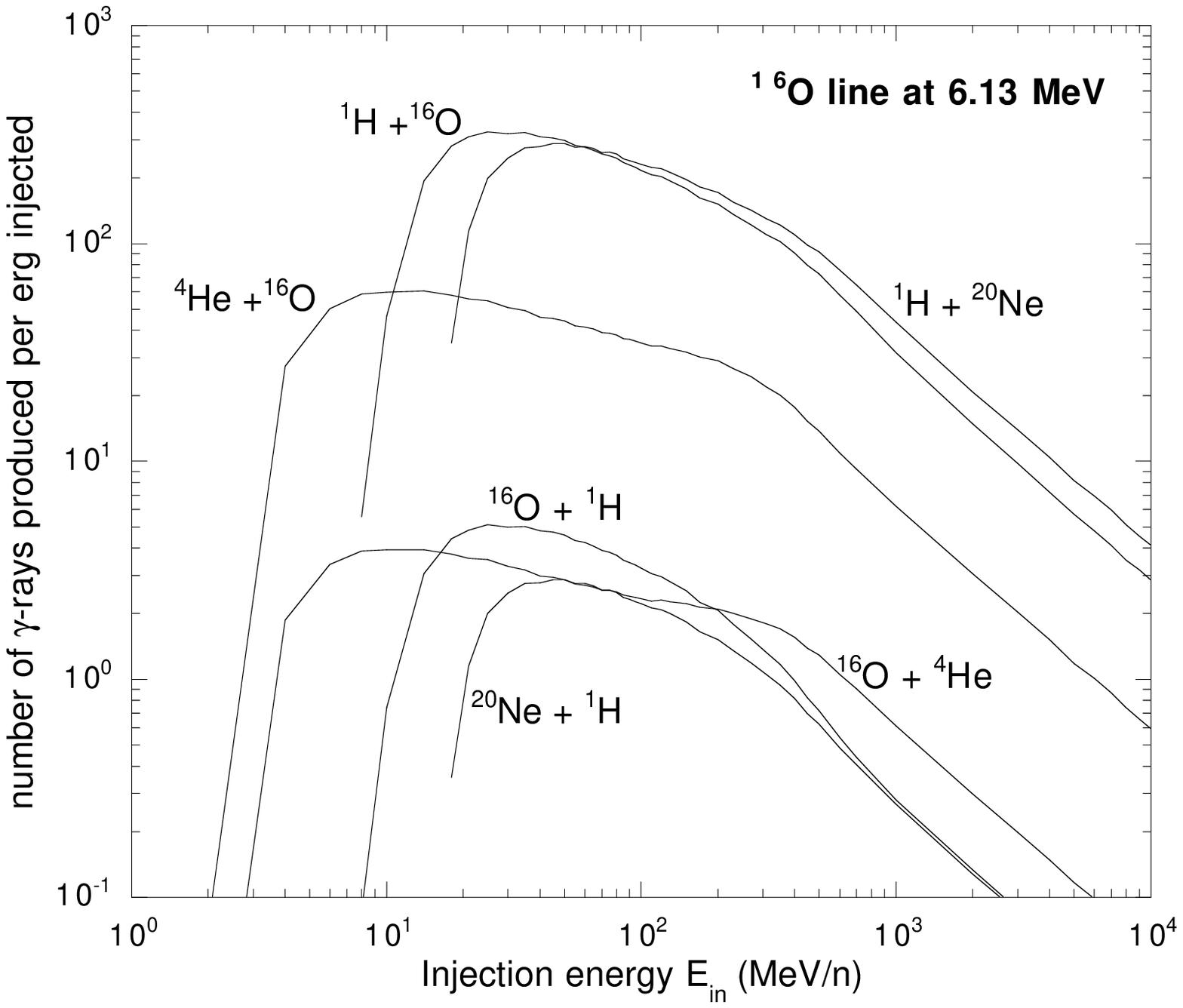}
	\caption{On the left (a): Absolute photon yields,
	$\mathcal{N}_{i,j;\gamma}$, in the 6.13~MeV line of $^{16}$O
	through various channels, as a function of the injection
	energy of the projectile.  The latter is the first nucleus
	appearing in the label, and the target is the second.  On the
	right (b): Gamma-ray production efficiency, in photon/erg, for
	the 6.13~MeV line of $^{16}$O through various channels, as a
	function of the injection energy of the projectile.}
	\label{fig:OLine}
    \end{figure*}

    \begin{figure*}
	\centering
	\includegraphics[width=8cm]{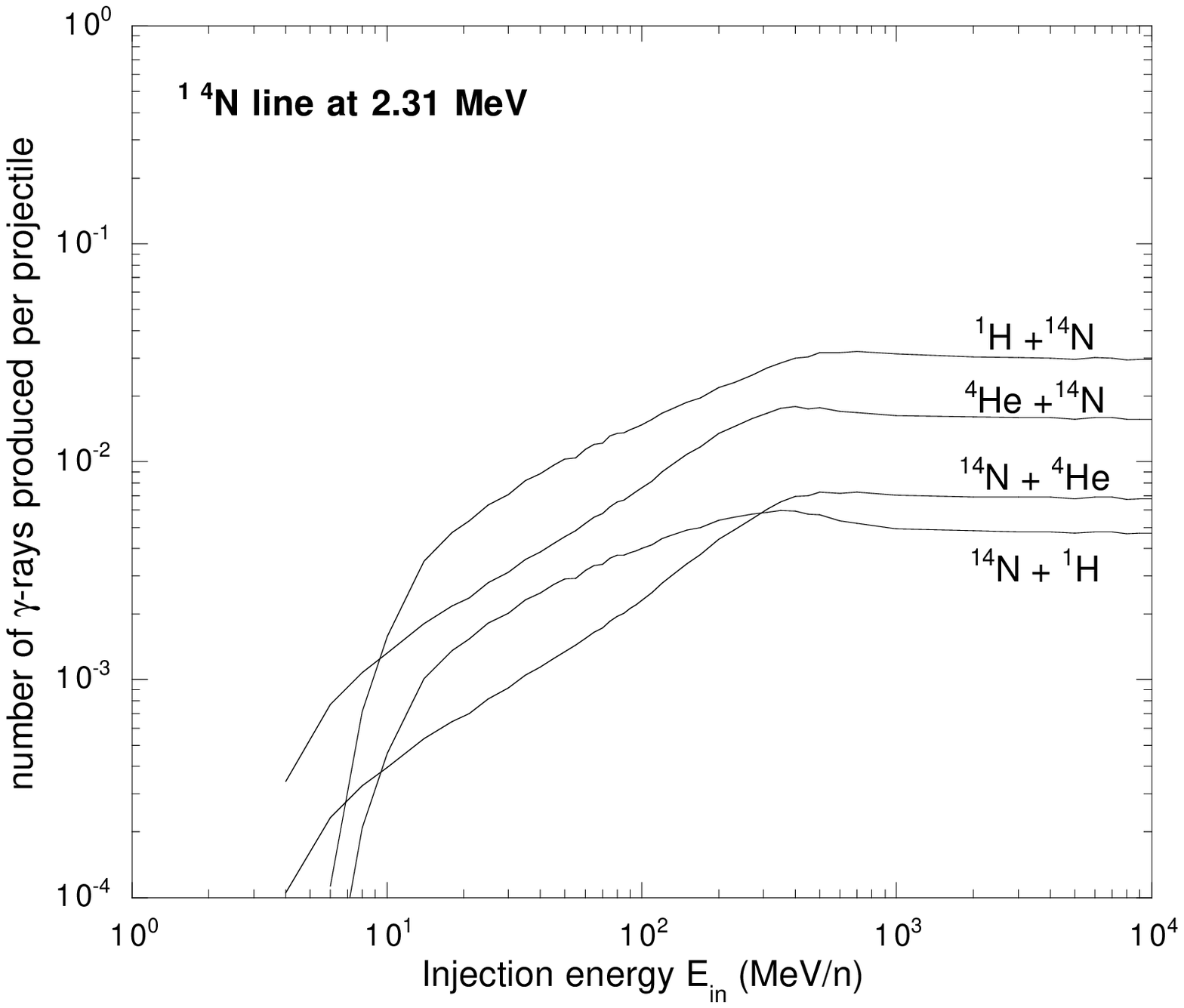}\hspace{0.2cm}
	\includegraphics[width=8cm]{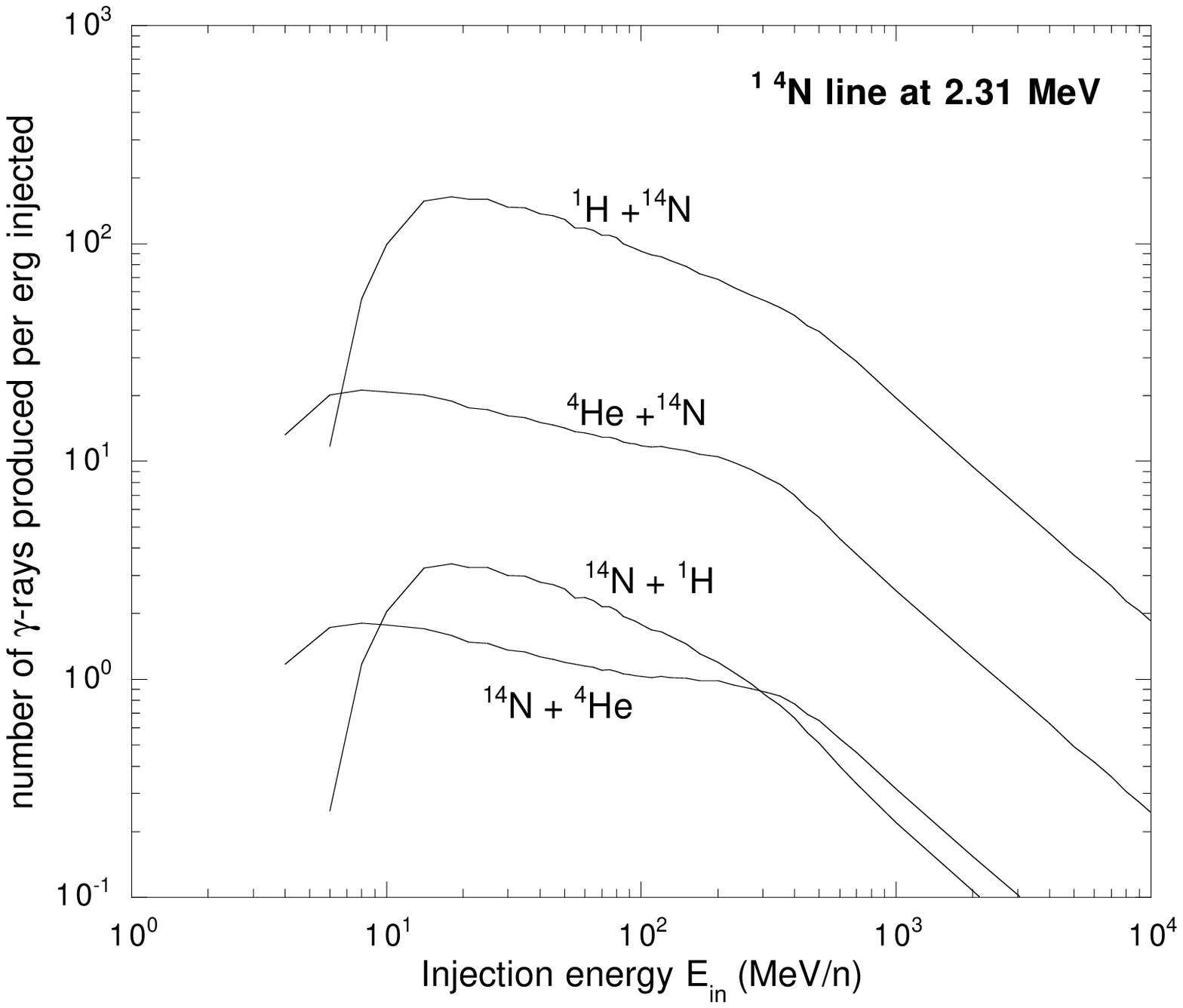}
	\caption{Same as Fig.~\ref{fig:OLine} for the $^{14}$N line at
	2.31~MeV.}
	\label{fig:NLine}
    \end{figure*}

    \begin{figure*}
	\centering
	\includegraphics[width=8cm]{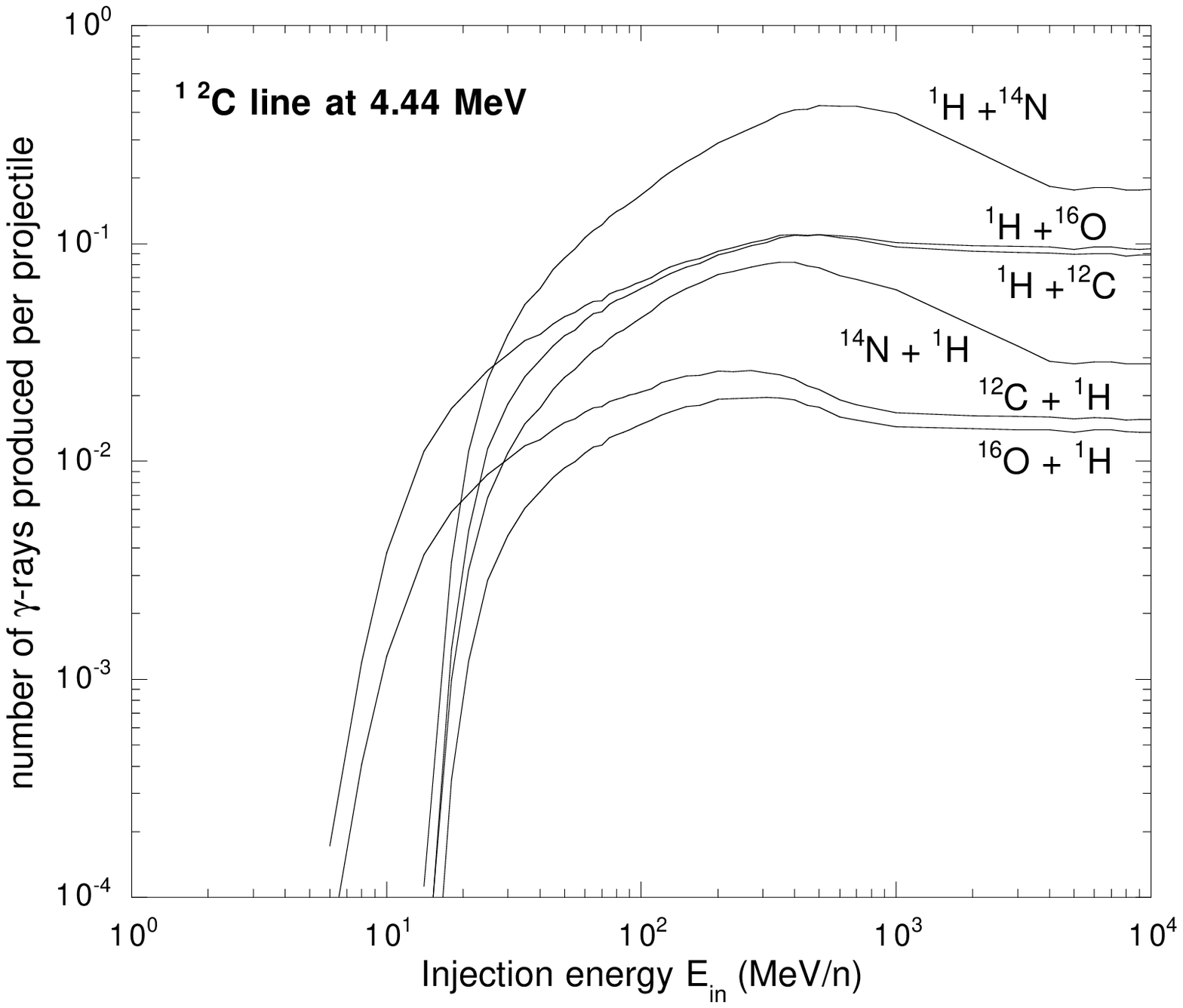}\hspace{0.2cm}
	\includegraphics[width=8cm]{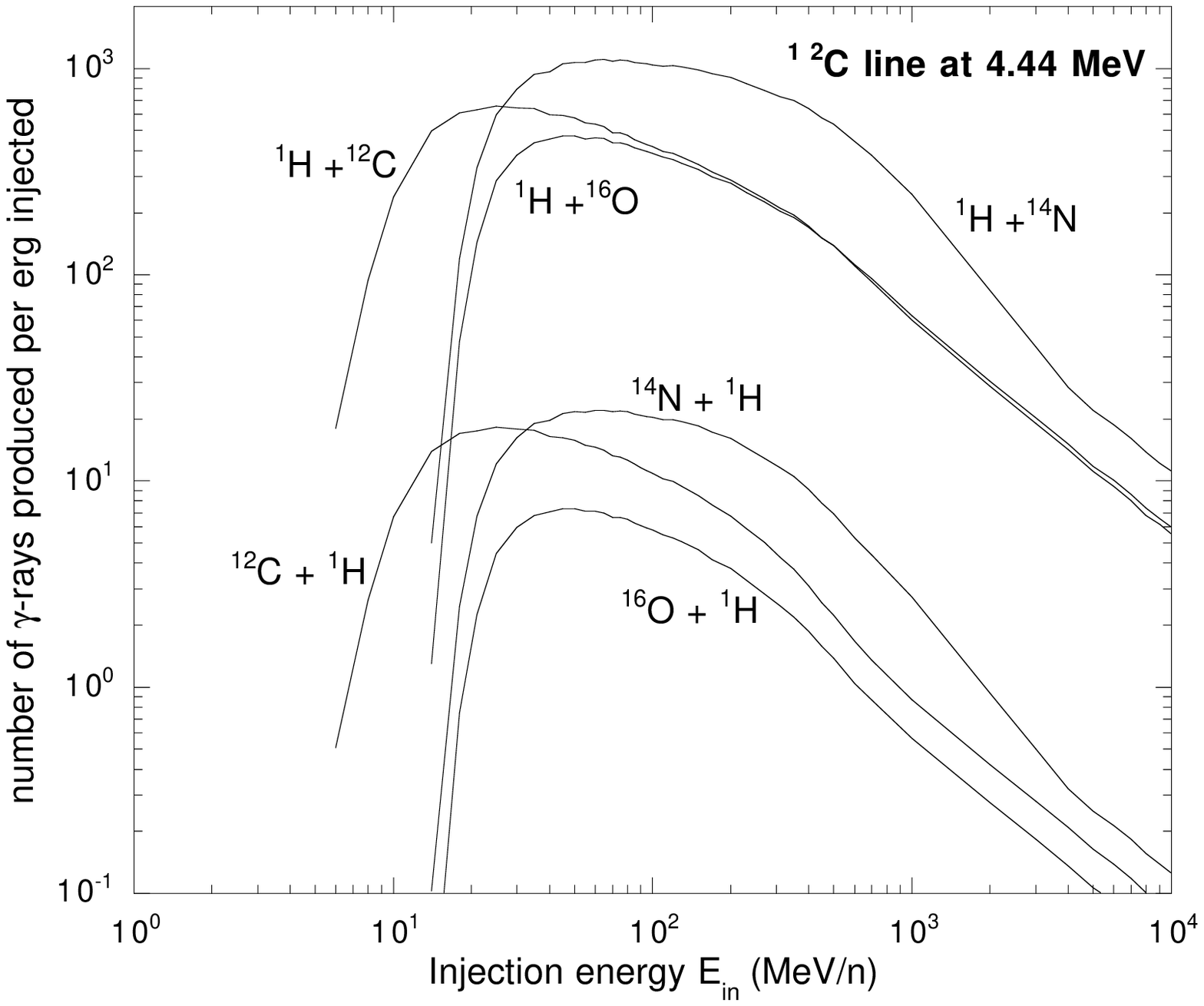}
	\\
	\includegraphics[width=8cm]{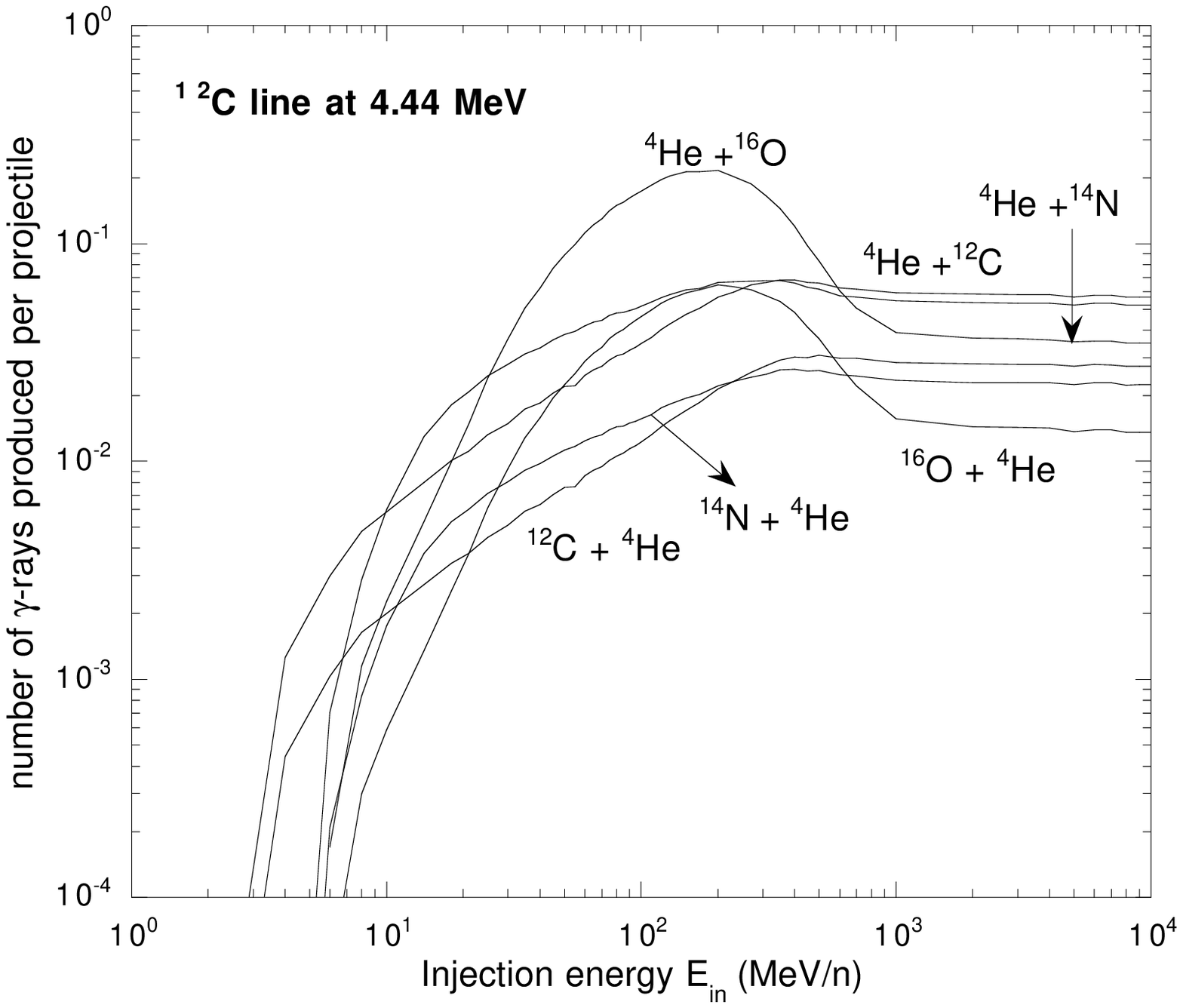}\hspace{0.2cm}
	\includegraphics[width=8cm]{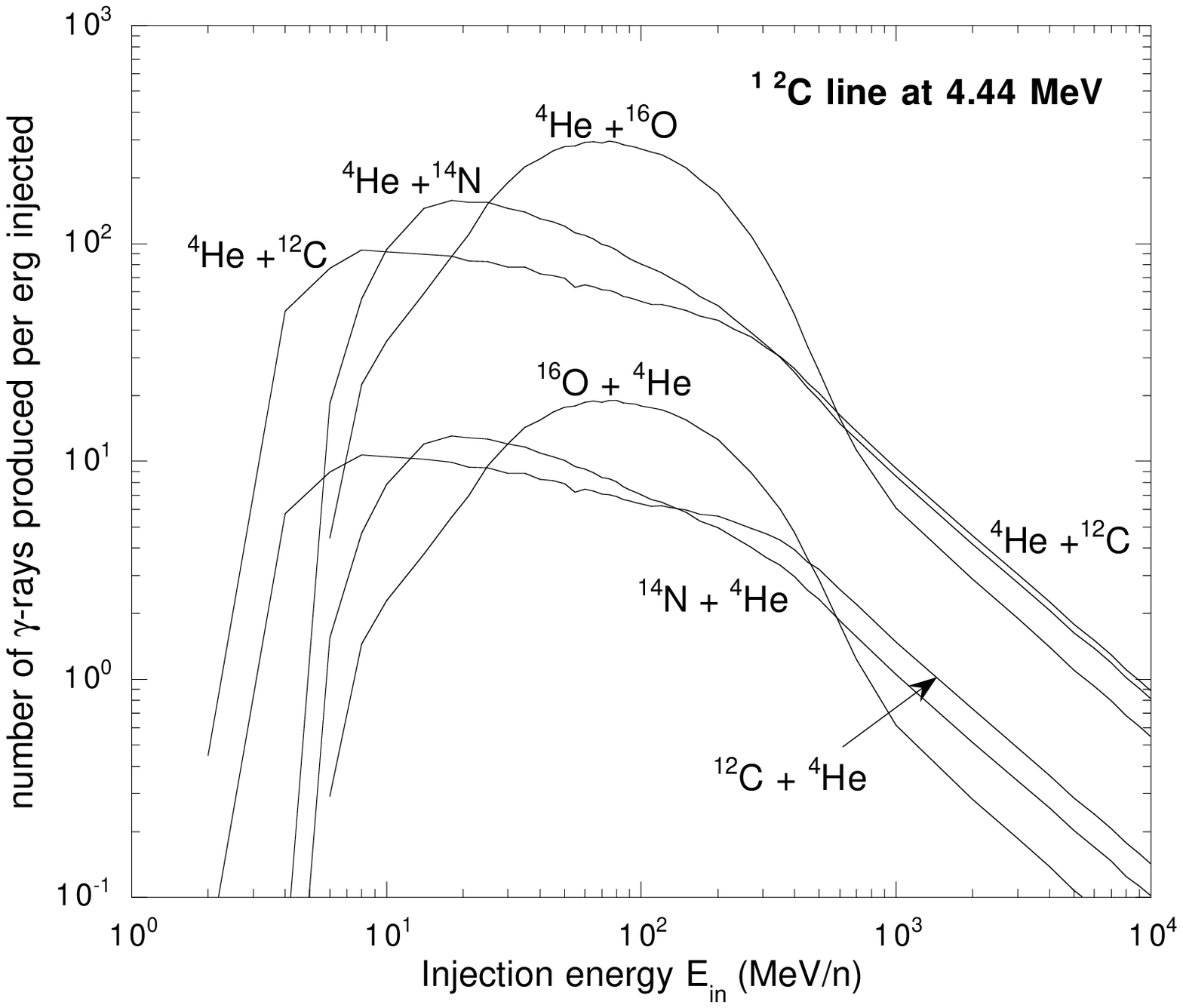}
	\caption{Same as Fig.~\ref{fig:OLine} for the $^{12}$C line at
	4.44~MeV, for reactions involving H nuclei (top) and He nuclei
	(bottom).}
	\label{fig:CLineH}
    \end{figure*}

    \begin{figure*}
	\centering
	\includegraphics[width=8cm]{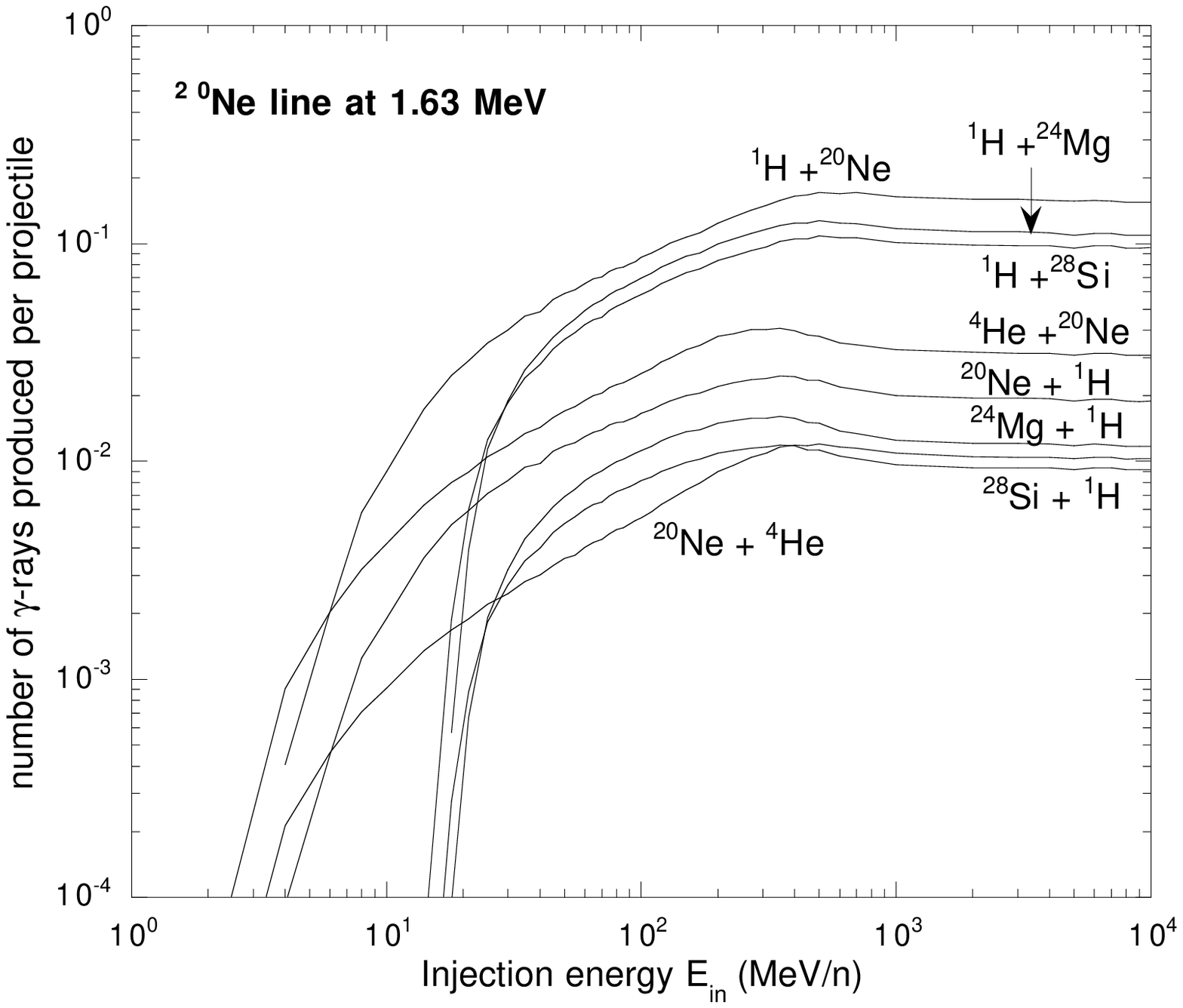}\hspace{0.2cm}
	\includegraphics[width=8cm]{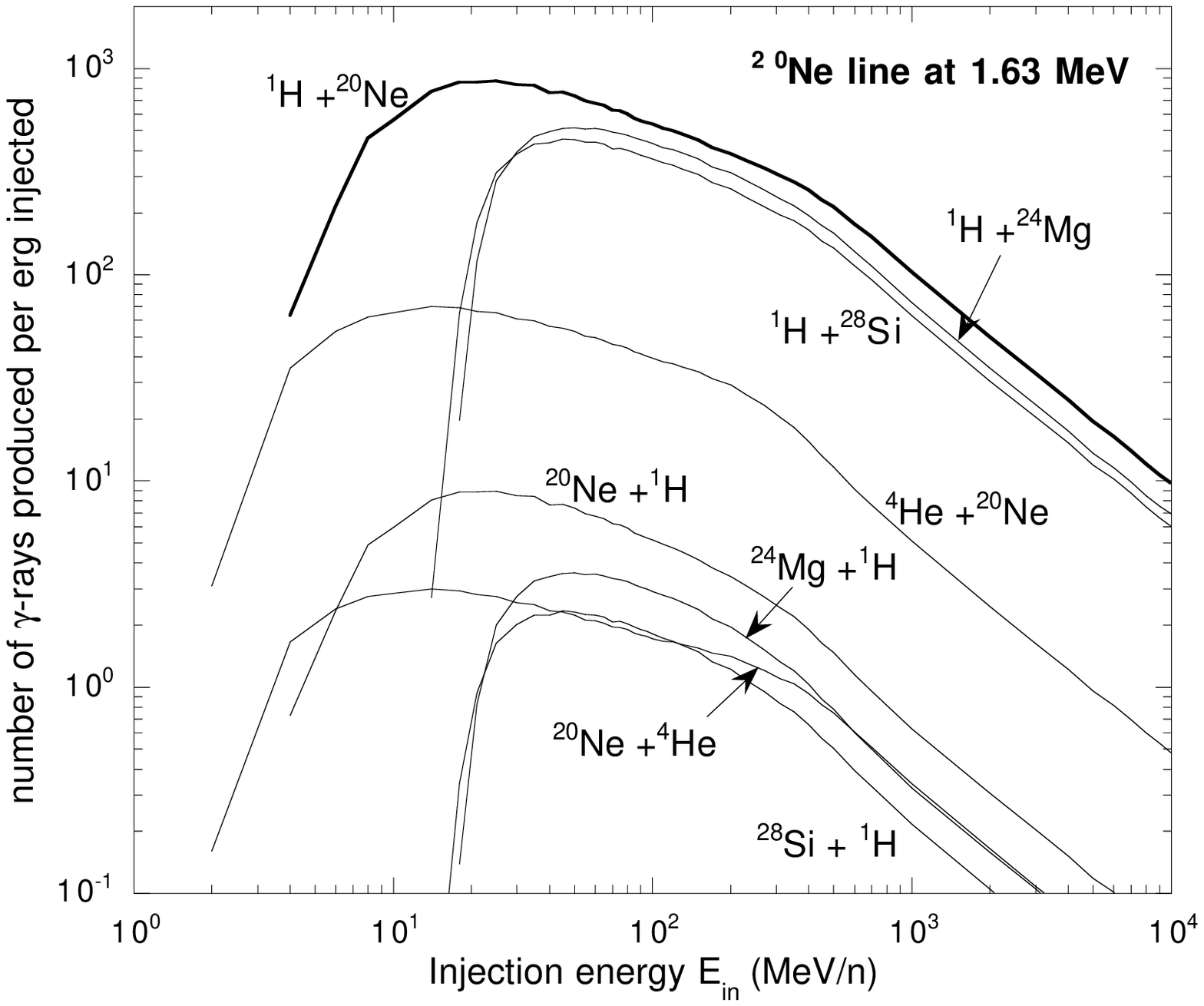}
	\\
	\includegraphics[width=8cm]{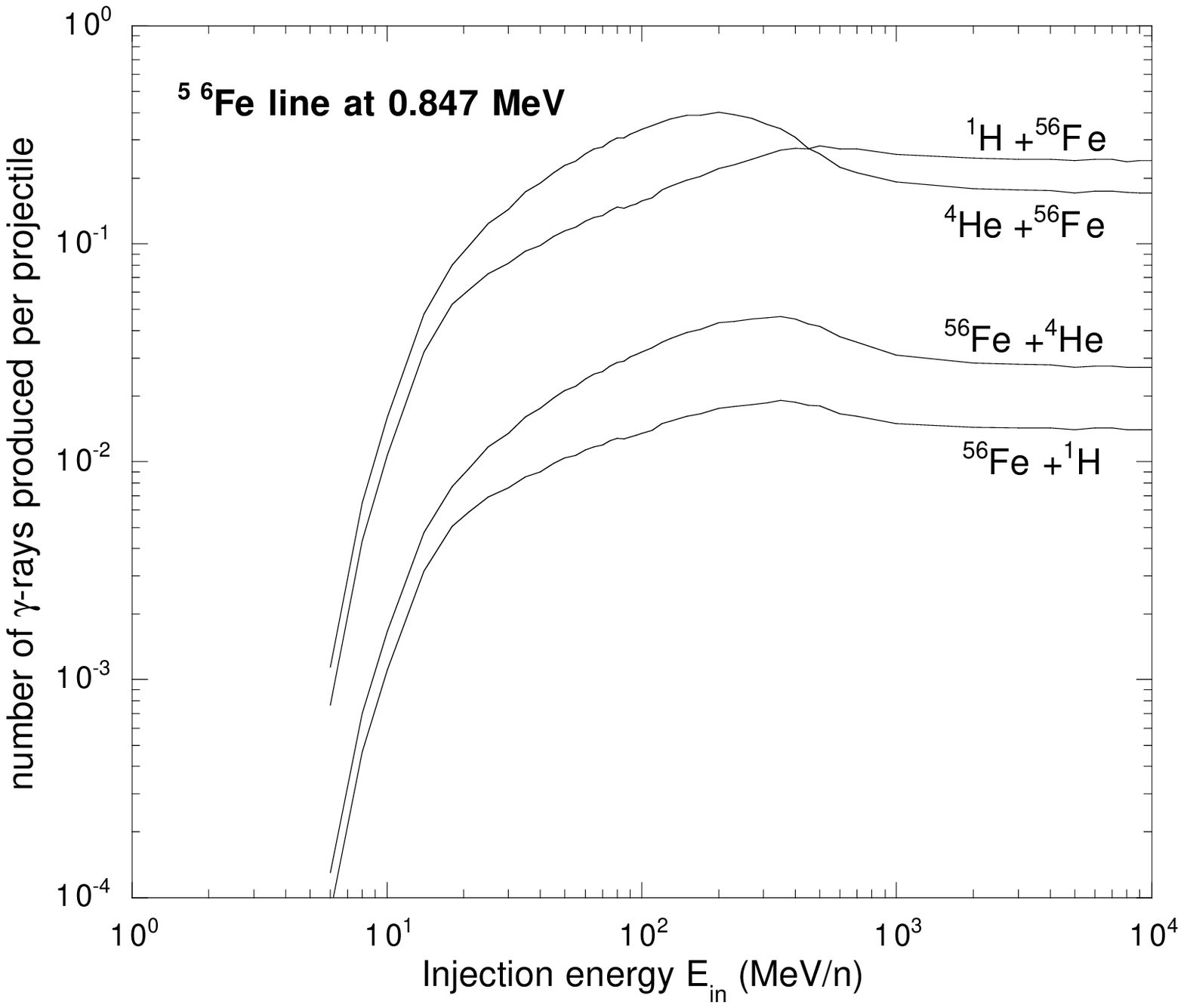}\hspace{0.2cm}
	\includegraphics[width=8cm]{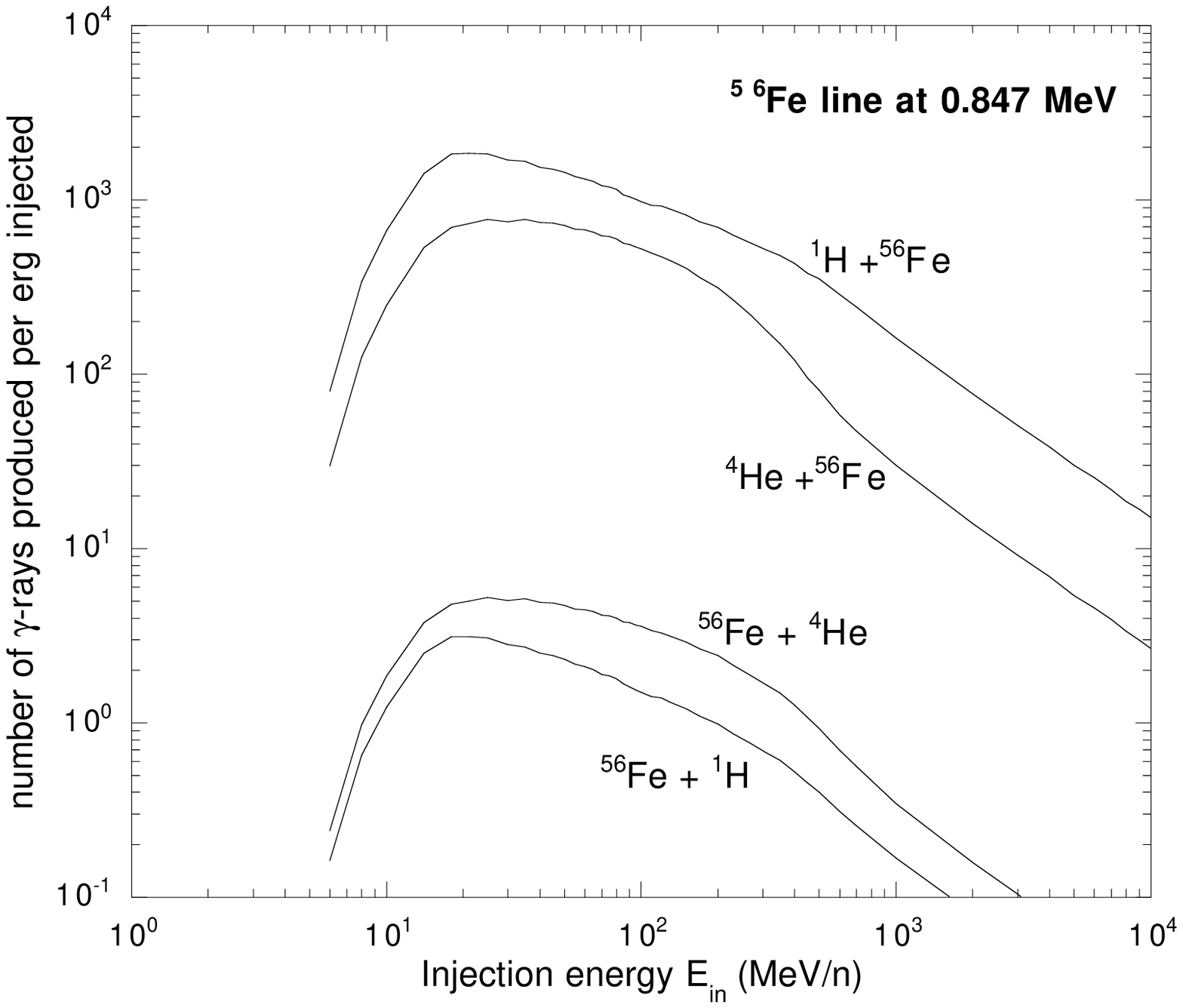}
	\caption{Same as Figs.~\ref{fig:OLine} for the $^{20}$Ne line
	at 1.63~MeV and the $^{56}$Fe line at 0.847~MeV.}
	\label{fig:NeAndFeLine}
    \end{figure*}

    The physical interpretation of
    $\mathcal{N}_{i,j;\gamma}(E_{\mathrm{in}})$ is straightforward: it
    is the number of photons of species $\gamma$ produced in a target
    made solely of nuclei of species $j$, by one projectile of species
    $i$ injected in the ISM at the energy $E_{\mathrm{in}}$,
    integrated over its entire life (i.e. from its injection until it
    has lost so much energy that it is below the nuclear excitation
    threshold).  Note that the lower bound of the integral can be
    replace by the energy threshold of the cross sections.  Now the
    interesting point is that the absolute photon yields,
    $\mathcal{N}_{i,j;\gamma}(E_{\mathrm{in}})$, can be calculated
    from physical quantities alone and is independent of astrophysics:
    as can be seen from Eqs.~(\ref{eq:PhotonYield})
    and~(\ref{eq:survivalProba}), it only depends on the nuclear
    cross sections and energy loss rates.  These can be calculated or
    measured once and for all, and so is it for
    $\mathcal{N}_{i,j;\gamma}(E_{\mathrm{in}})$.

    As anticipated, the great advantage of this formulation is that
    once the quantities $\mathcal{N}_{i,j;\gamma}(E_{\mathrm{in}})$
    have been calculated, the actual $\gamma$-ray emission rate in a
    given astrophysical situation can be derived from
    Eq.~(\ref{eq:ProdRateInverted}) which gathers all the
    astrophysical information (namely the EP spectrum and composition,
    and the target composition), but which is now expressed in terms
    of the \textit{source spectrum}, rather than the
    \textit{propagated} one.  To better understand the signification
    of this transformation, it suffices to compare
    Eqs.~(\ref{eq:TauxProdStat}) and~(\ref{eq:ProdRateInverted}).  We
    have replaced the propagated spectral density of the EPs,
    $N_{i}(E)$, by their injection function, $Q_{i}(E)$, and the
    cross sections $\sigma_{i,j;\gamma}$ by our absolute photon
    yields, $\mathcal{N}_{i,j;\gamma}$, which play the role of
    `effective cross-sections' (although their physical dimension is
    different) taking into account the propagation of the EPs in the
    ambient medium.  It should be stressed that the individual photon
    yields behave as universal physical quantities and can be used
    with any source spectrum, any EP composition \textit{and} any
    target composition.

    Two comments are in order here.  First, the above expression
    giving the photon yields
    $\mathcal{N}_{i,j;\gamma}(E_{\mathrm{in}})$ may seem to depend on
    the density, $n_{0}$, of the propagation medium (e.g. the ISM). 
    This is actually not the case, as the energy loss rate appearing
    in the denominator is also proportional to this density.  The
    second comment concerns the universality of EP propagation, which
    is crucial in the approach developed here.  Indeed, in principle
    the EP energy loss rates and survival probabilities depend on the
    propagation medium, so that a different photon yield should be
    calculated for each propagation medium.  These specific photon
    yields could still be used with any source spectrum and
    composition, but not with any target composition, as the latter is
    usually the same as that of the propagation medium.  However, as
    shown in Sect.~\ref{sec.universality}, the dependence of the
    energy loss rates and survival probabilities with metallicity is
    negligible in most situations, so that the photon yields
    $\mathcal{N}_{i,j;\gamma}$ can indeed be considered universal.

    \section{Results and emission rates reconstruction}
    \label{sec:Results}

    \subsection{Nuclear excitation cross sections}
    \label{Sect:CrossSections}
   
    Before showing the individual photon yields calculated for the
    main expected gamma-ray lines in the ISM, we should say a word
    about cross sections, as they provide the major uncertainty in our
    results and most of them are extrapolated from relatively scarce
    experimental data.  We have used the data from Ramaty et al. 
    (1979), updated with more recent experimental data whenever
    possible (Dyer et al., 1981,1985; Lang et al., 1987; Lesko et al. 
    1988; Kiener et al., 2001).  The paper by Kiener et al.  (2001,
    and private communication) has been used to derive a general
    procedure to extrapolate the data at high energy.
   
    Above the resonance peak of the cross sections, whenever the data
    is missing we assume that the cross section obeys a simple law in
    $a\times E^{-x} + b$ (above 20--30~MeV/n, say).  The best fit of
    the data for the $^{12}$C(p,p$\gamma$) reaction gives $\sigma(E) =
    24337 \times E^{-1.74} + 4$~mbarn, where $E$ is in MeV. For the
    $^{12}\mathrm{C}(\alpha,\alpha\gamma)$ reaction, one finds
    $\sigma(E) = 31400 \times E^{-1.51} + 7$~mb, and for the
    excitative spallation reaction
    $^{16}\mathrm{O(p,p}\alpha)^{12}\mathrm{C}^{*}_{4.44}$, $\sigma(E)
    = 75951 \times E^{-1.98} + 4.3$~mb (Kiener, 2001; private
    communication).
   
    \begin{table}
	\caption[]{Data relative to the fit of the cross section as
	given by Eq.~\ref{eq:sigmaAbovePeak}.  The reaction is given
	in the first column, with the names of the target and
	projectile and the energy of the resulting gamma ray in MeV.
	The second and third columns give the energy of the cross
	section peak and the corresponding value of the cross
	section.}
	\label{tab:crossSectionsData}
	\begin{tabular}{lcc} 
	    \hline
	    \hspace{17pt}reaction & $E_{\mathrm{peak}}$ & $\sigma_{\mathrm{peak}}$ \\
	    \hline
	    $^{1}$H+$^{56}$Fe=0.847  & 13.0 & 861 \\
	    $^{4}$He+$^{56}$Fe=0.847 & 13.0 & 1290 \\
	    $^{1}$H+$^{20}$Ne=1.634  & 11.0 & 330 \\
	    $^{4}$He+$^{20}$Ne=1.634 & 3.19 & 232 \\
	    $^{1}$H+$^{24}$Mg=1.634  & 23.0 & 180 \\
	    $^{1}$H+$^{28}$Si=1.634  & 21.7 & 157 \\
	    $^{1}$H+$^{14}$N=2.313   & 10.0 & 106 \\
	    $^{4}$He+$^{14}$N=2.313  & 3.25 & 119 \\
	    $^{1}$H+$^{12}$C=4.438   & 11.0 & 317 \\
	    $^{4}$He+$^{12}$C=4.438  & 3.00 & 393 \\
	    $^{1}$H+$^{14}$N=4.438   & 21.7 & 291 \\
	    $^{4}$He+$^{14}$N=4.438  & 9.64 & 333 \\
	    $^{1}$H+$^{16}$O=4.438   & 22.5 & 156 \\
	    $^{4}$He+$^{16}$O=4.438  & 25.0 & 223 \\
	    $^{1}$H+$^{16}$O=6.129   & 13.0 & 163 \\
	    $^{4}$He+$^{16}$O=6.129  & 3.50 & 269 \\
	    $^{1}$H+$^{20}$Ne=6.129  & 19.0 & 108 \\
	    \hline
	\end{tabular}
    \end{table}
    
    As a first approximation, we assume that the excitation cross
    sections of other nuclei (namely, $^{14}$N, $^{16}$O, $^{20}$Ne
    and $^{56}$Fe) obey the similar laws with the same value of the
    exponent $x$, i.e. 1.74 and 1.51 respectively for proton-induced
    and alpha-particle-induced excitations.  The constant value at
    high energy, $B$, is simply taken as the value of $B$ for the
    $^{12}$C excitation cross sections, but scaled to the measured
    value of the cross section at the resonance peak (i.e.
    proportionally to the peak value).  Finally, the value of $A$ is
    obtained by imposing the continuity of the cross sections above
    the peak.  The same procedure is applied to excitative spallation
    reactions (i.e. an exponent of 1.98 and a high energy value scaled
    proportionally to the peak value), although such an extrapolation
    is more problematic in this case, as the $^{16}$O nucleus has an
    atypical structure with a large component of four $\alpha$
    particles.

    The above procedure can be summarized by the following expression
    giving the cross section above the peak (at energy
    $E_{\mathrm{peak}}$ and with cross section
    $\sigma_{\mathrm{peak}}$), in terms of the values of the reference
    cross section at the peak, $\sigma_{\mathrm{peak}}^{0}$, and at
    high energy, $\sigma_{\infty}^{0}$, given above:
    
    \begin{equation}
	\sigma(E) = \sigma_{\mathrm{peak}}\left[
	\left(1 - \frac{\sigma_{\infty}^{0}}{\sigma_{\mathrm{peak}}^{0}}
	\right)
	\left(\frac{E}{E_{\mathrm{peak}}}\right)^{-x}
	+ \frac{\sigma_{\infty}^{0}}{\sigma_{\mathrm{peak}}^{0}}
	\right]
	\label{eq:sigmaAbovePeak}
    \end{equation}
    
    For completeness, we give the values of $E_{\mathrm{peak}}$ and
    $\sigma_{\mathrm{peak}}$ for the various cross sections considered
    in this paper in Table~\ref{tab:crossSectionsData}.
    
    In general, the error on the excitation cross sections and thus on
    the gamma-ray yields is typically 10\% whenever actual
    experimental data exist (this is the case for the two main lines
    of $^{12}$C and $^{16}$O), and of the order of 20\% to 50\% when
    the values are simply estimated or extrapolated.  This is quite
    substantial, especially when our goal is to look at line ratios,
    in the hope to determine the composition of the EPs and/or the
    ambient medium from gamma-ray line measurements.  These errors,
    unfortunately, cannot be lowered but by increasing the
    experimental effort at terrestrial accelerators.  This is strongly
    recommended in order to make the most of the opening field of
    gamma-ray astronomy.
   
    \subsection{Photon yields for the $^{12}$C, $^{14}$N, $^{16}$O,
    $^{20}$Ne and $^{56}$Fe $\gamma$-ray lines.}
   
    In Fig.~\ref{fig:OLine}a, we show the absolute $\gamma$-ray
    yields, $\mathcal{N}_{i,j;\gamma}$, corresponding to the main
    $^{16}$O line at 6.13~MeV, for various projectiles and targets.

    The evolution of $\mathcal{N}_{i,j;\gamma}$ as the injection
    energy increases can be interpreted in the following way.  Photon
    production begins when $E_{\mathrm{in}}$ becomes greater than the
    reaction threshold.  Then it increases sharply as
    $E_{\mathrm{in}}$ passes through the peak of the cross section,
    and increases more smoothly afterwards.  As long as particle
    destruction or escape can be neglected, Eq.~(\ref{eq:PhotonYield})
    makes it clear that the number of photons produced is an
    increasing function of $E_{\mathrm{in}}$, the upper bound of the
    integral.  Physically, the particle produces $\gamma$-rays all the
    way as its energy goes down from $E_{\mathrm{in}}$ to below the
    reaction threshold.  If it is injected at a higher energy, it will
    produce $\gamma$-rays for a longer time, integrating the cross
    section over a larger range of energy.
   
    But when $E_{\mathrm{in}}$ increases further, there comes a time
    when the projectile has a large probability of being destroyed
    (through a nuclear reaction) or escaping from the region under
    study (in a thin target model), \textit{before} its energy drops
    below the reaction threshold.  In this case, the effective energy
    range over which the cross section is integrated is reduced from
    below, and the overall $\gamma$-ray yield starts to decrease.  For
    large enough $E_{\mathrm{in}}$, the particle never reaches the
    most efficient energy range corresponding to the peak of the cross
    section.  Since both the destruction and the excitation cross
    sections are roughly constant at high energy, the photon yield
    tends to an `asymptotic value' where increasing the injection
    energy only shifts upwards the energy range of activity of the EP
    but does not change the integrated photon yield.  This asymptotic
    value merely depends on the ratio of the excitation and
    destruction cross sections.
   
    Typically, for the main $\gamma$-ray lines to be expected in the
    ISM, one can see from Figs.~\ref{fig:OLine}
    to~\ref{fig:NeAndFeLine} that several percent to up to 30\% of the
    projectiles injected will produce a gamma-ray, with a peak of this
    number in the range $E_{\mathrm{in}} = 100$--300~MeV/n.

    While the decrease of $\mathcal{N}_{i,j;\gamma}$ at high energy is
    not very steep, it should be realized that the $\gamma$-ray
    production efficiency, defined as the number of photons produced
    per erg of projectiles injected, is falling down more quickly, as
    shown on the right sides of Figs.~\ref{fig:OLine}
    to~\ref{fig:NeAndFeLine}, for the same reactions as on the left
    sides.  The corresponding curves give a visual representation of
    the most efficient energy range for an EP to produce a given
    $\gamma$-ray line.  They can be thought of as simple
    phenomenological tools: a simple look at them gives an idea of the
    kind of source spectrum and composition required to reproduce any
    $\gamma$-ray line observational data.  Note that contrary to what
    might have been naively expected, this range starts at an energy
    higher than the cross section peak, and extends to even higher
    energies.  In other words, the most efficient way to produce
    gamma-rays in the ISM is to use EPs with energies between, say, 10
    to 300~MeV/n.

    \subsection{Gamma-ray line emission synthesis}
   
    \begin{figure*}
	\centering
	\includegraphics[width=8cm]{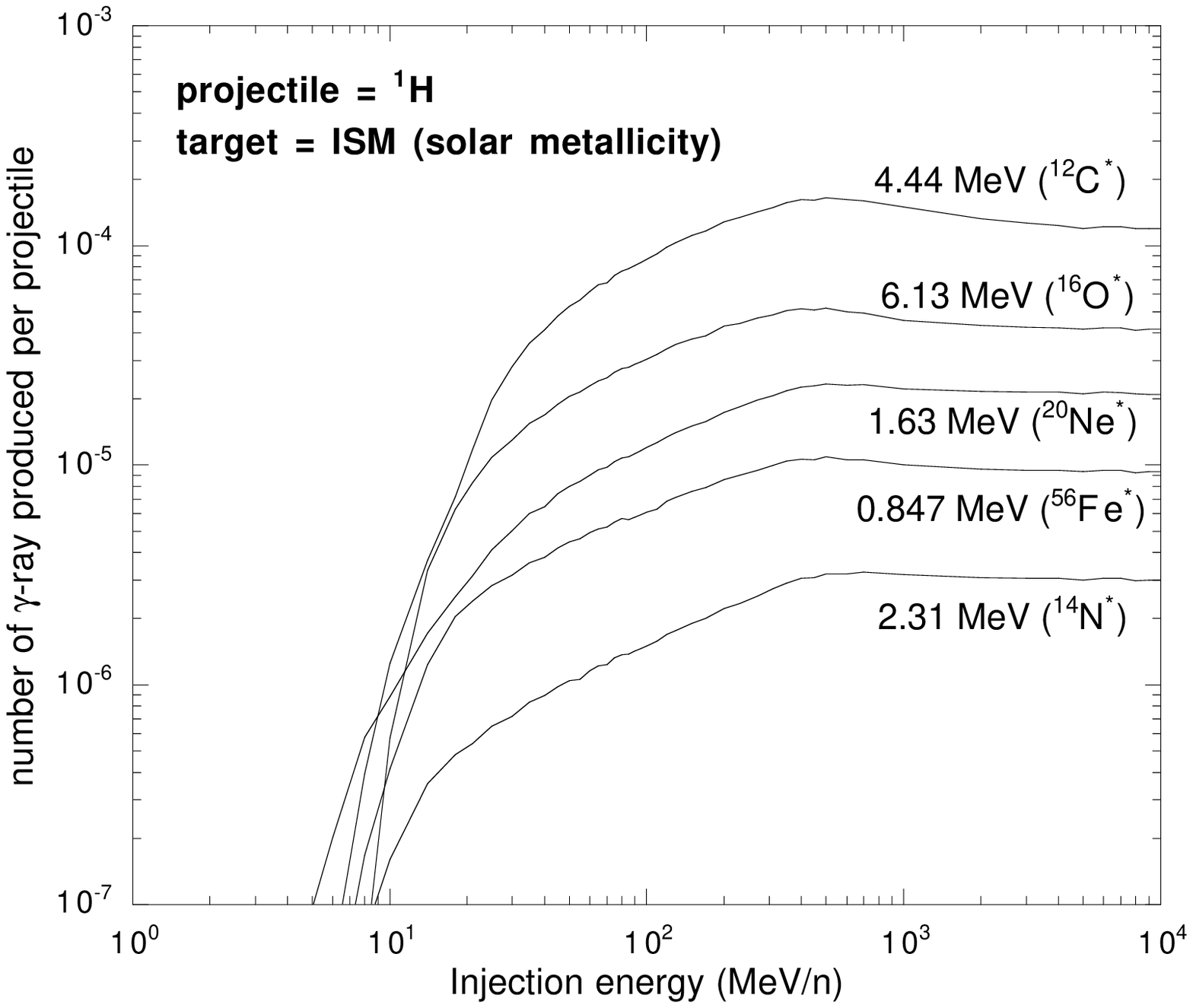}\hspace{0.2cm}
	\includegraphics[width=8cm]{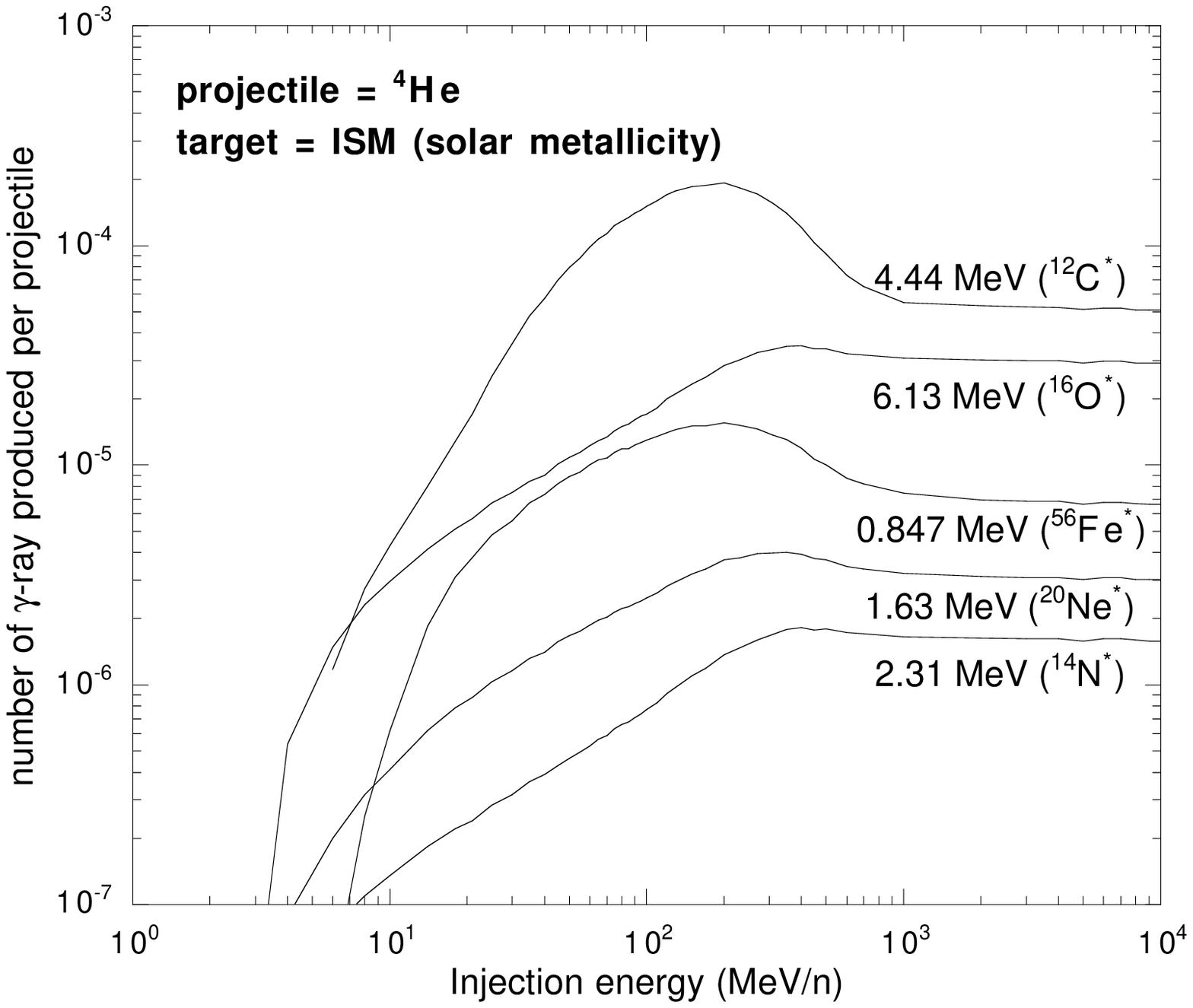}
	\caption{Gamma-ray yields of a $^{1}$H (left) and a $^{4}$He
	(right) nucleus injected in a medium of solar metallicity, as
	a function of the injection energy.  Contributions to various
	$\gamma$-ray lines are shown.}
	\label{fig:HAndHeInISM}
    \end{figure*}

    From a practical point of view, the quantities
    $\mathcal{N}_{i,j;\gamma}(E_{\mathrm{in}})$ also allow one to
    straightforwardly calculate the $\gamma$-ray line emission in a
    given astrophysical situation.  It suffices to sum the
    contributions of each reaction involved, weighted according to the
    desired chemical abundances of both the source and the target.  In
    other words, one can calculate the $\gamma$-ray line emission rate
    for \textit{any} EP spectrum and composition in \textit{any}
    medium (except maybe the most extremely metal-rich), without
    needing to worry about particle propagation and energy losses at
    all, as intended.

\begin{figure}
	\vspace*{2.0mm}
	\includegraphics[width=8.3cm]{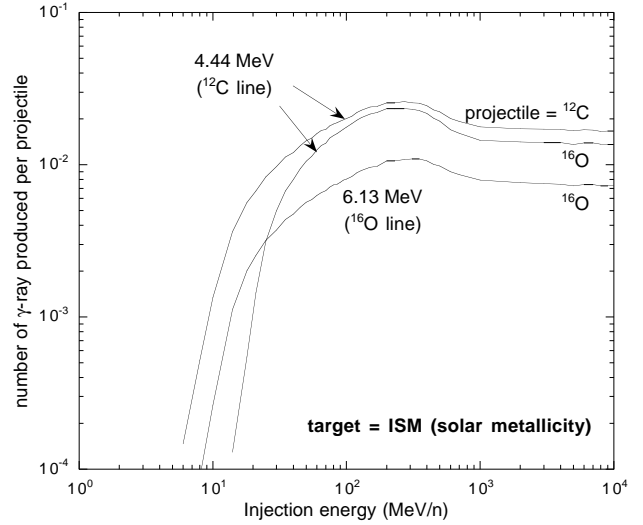}
	\caption{Same as Fig.~\ref{fig:HAndHeInISM} for $^{12}$C and
	$^{16}$O projectiles.}
	\label{fig:CetOInISM}
    \end{figure}

    Such a weighting is illustrated in Figs.~\ref{fig:HAndHeInISM}
    and~\ref{fig:CetOInISM}, where we show the $\gamma$-ray yields of
    H, He, C and O nuclei in a medium of solar metallicity.  Note that
    although the $\gamma$-ray yields of C and O projectiles appear
    much higher than those of He (or H), they still have to be
    weighted by the relative abundances of the various projectiles
    among the EPs.  The number of gamma-rays in the C and O lines
    produced by one EP injected in the ISM at energies above a few
    tens of MeV/n is typically between $10^{-5}$ and $10^{-4}$.
   
\begin{figure}
	\vspace*{2.0mm}
	\includegraphics[width=8.3cm]{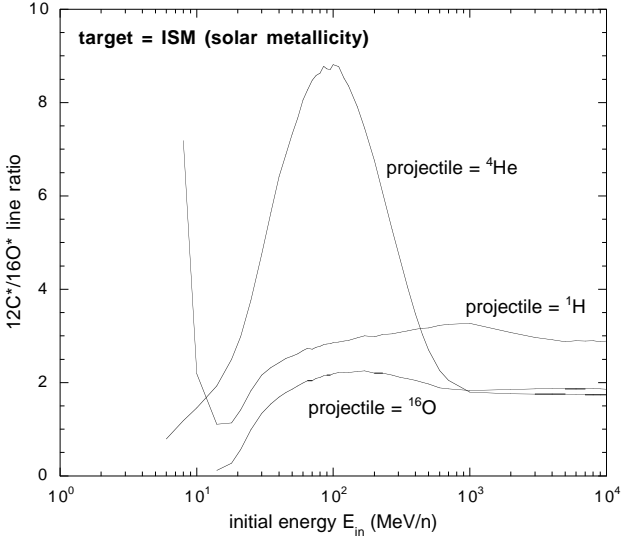}
	\caption{$^{12}\mathrm{C}^{*}$/$^{16}\mathrm{O}^{*}$ emission
	line ratio produced by H, He and O nuclei injected into the ISM
	with solar metallicity, as a function of the injection energy.}
	\label{fig:COLineRatio}
    \end{figure}

    An interesting result is the fact that, in addition to $^{12}$C
    nuclei, $^{16}$O nuclei are also rather efficient in producing the
    $^{12}$C line at 4.44~MeV. In Fig.~\ref{fig:COLineRatio}, we have
    also shown the $^{12}\mathrm{C}^{*}$/$^{16}\mathrm{O}^{*}$
    emission line ratio for the three projectiles producing both of
    these lines, namely H, He and O. This can be used to estimate
    quickly the probable composition of EPs producing any observed
    $^{12}\mathrm{C}^{*}$/$^{16}\mathrm{O}^{*}$ line ratio.

    \section{Analytical estimates}
   
    In an earlier work, Bykov and Bloemen (1994) calculated the
    average photon yield of a single energetic nucleus suffering from
    Coulombian energy losses (neglecting nuclear destruction).  It is
    worth comparing our results with their analytical approximation,
    obtained with the Bethe-Bloch formula for energy losses, and
    extending their approached formula to situations where the
    excitation cross section follows a simple analytical law (such as
    that mentioned in Sect.~\ref{Sect:CrossSections}), and where the
    nuclear destruction is important.
    
    \begin{figure}[t]
	\centering
	\includegraphics[width=8cm]{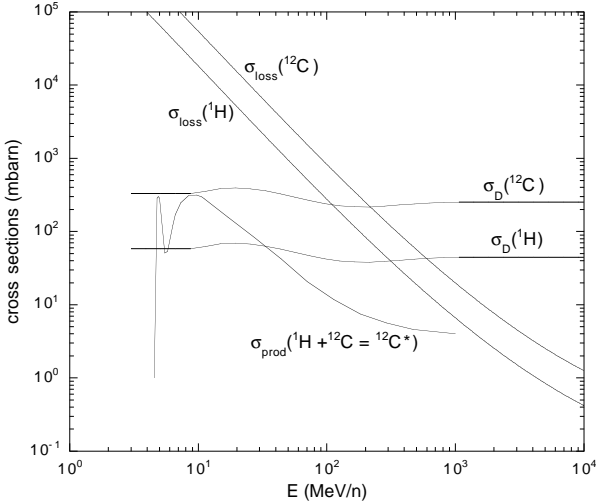}
	\caption{Comparison of the total inelastic cross section
	(nuclear destruction) and the energy loss cross section for
	$^{12}$C and $^{1}H$.  The $^{12}$C excitation cross section
	is also shown for comparison.}
	\label{fig:lossVsDestruction}
    \end{figure}

    Depending on the EP injection energy, one can identify two
    opposite `regimes' where the estimation of the photon yields
    $\mathcal{N}_{i,j;\gamma}(E_{\mathrm{in}})$ can be simplified. 
    From the physical point of view, this depends whether the particle
    energy losses can be neglected with respect to nuclear
    destruction, or vice versa.  On
    Fig.~\ref{fig:lossVsDestruction}, we have drawn the destruction
    cross section, $\sigma_{\mathrm{D}}(E)$, together with the energy
    loss cross section, $\sigma_{\mathrm{loss}}(E)$, defined by:
    \begin{equation}
	\sigma_{\mathrm{loss}}(E) =
	\frac{1}{n_{0}v(E)\tau_{\mathrm{loss}}(E)} =
	\frac{|\dot{E}(E)|}{nvE},
	\label{eq:sigmaLoss}
    \end{equation}
    for energetic $^{1}$H and $^{12}$C nuclei. Note that with this 
    definition, Eq.~(\ref{eq:PhotonYield}) can be rewritten in a 
    simple way as:
    \begin{equation}
	\mathcal{N}_{\gamma}(E_{\mathrm{in}}) =
	\int_{0}^{E_{\mathrm{in}}}
	\frac{\sigma_{\mathrm{prod}}(E)}{\sigma_{\mathrm{loss}}(E)}
	\,\mathcal{P}_{i}(E_{\mathrm{in}},E) \frac{\d E}{E}.
	\label{eq:PhotonYield2}
    \end{equation}
    
    It can be seen on Fig.~\ref{fig:lossVsDestruction} that at low
    energy, destruction is negligible compared to energy losses, while
    the opposite is true at high energy.  The exact transition energy
    depends on the nucleus considered, and ranges between 200 and
    300~MeV/n.  This is consistent with the results of
    Fig.~\ref{fig:survivalProba}, showing the transition of the
    survival probability from 1 to 0 around this energy.

    \subsection{Low energy limit}
    \label{sec:LowEnergyLimit}

    In the case when particle destruction can be neglected, one can
    set the survival probability,
    $\mathcal{P}_{i}(E_{\mathrm{in}},E)$, equal to 1 in
    Eq.~(\ref{eq:PhotonYield2}) (this is the case studied by Bykov and
    Bloemen, 1994).  The Bethe-Bloch formula for the energy losses of
    a nucleus (Z,A) gives:
    \begin{equation}
	\sigma_{\mathrm{loss}}(E) = \frac{Z^{2}}{A}\times
	\frac{3}{2}\sigma_{\mathrm{T}}\ln(\Lambda)\times
	\frac{m_{\mathrm{e}}c^{2}}{E}\times\frac{c^{2}}{v^{2}},
	\label{eq:Bethe-Bloch}
    \end{equation}
    where $\sigma_{\mathrm{T}}$ is the Thomson cross section and $\ln
    \Lambda$ is the usual Coulomb logarithm (see e.g. Lang, 1999).  It
    is a slightly increasing function of energy ($\sim 7.3$ at 
    10~MeV/n, and $\sim 12$ at 1~GeV/n).
    
    Reporting in Eq.~(\ref{eq:PhotonYield2}) and assuming that the
    excitation cross section is constant, one obtains the result of
    Bykov and Bloemen (1994) in the non-relativistic limit ($E \sim
    m_{\mathrm{p}}v^{2}/2$):
    \begin{equation}
	\mathcal{N}_{\gamma}(E_{\mathrm{in}}) =
	\frac{A}{Z^{2}}\frac{m_{\mathrm{p}}}{6m_{\mathrm{e}}\ln\Lambda}\,
	\frac{\sigma_{\mathrm{prod}}}{\sigma_{\mathrm{T}}}
	\left(\frac{v}{c}\right)^{4},
	\label{eq:BykBlo94}
    \end{equation}
    where the cross section can then be considered as a function of
    energy, although this amounts to moving
    $\sigma_{\mathrm{prod}}(E)$ outside the integral in
    Eq.~(\ref{eq:PhotonYield2}), which is of course improper in
    principle.
    
    The above formula can be improved by using an approached
    expression for the nuclear excitation cross-section, which allows
    an analytical integration of Eq.~\ref{eq:PhotonYield2} (with
    $\mathcal{P}_{i}(E_{\mathrm{in}},E) = 1$).  By approximating the
    cross sections as $\sigma_{\mathrm{prod}}(E) = a\times E^{-x} + b$
    above the peak (the values for $a$ and $b$ being deduced from
    Eq.~(\ref{eq:sigmaAbovePeak}) and
    Table~\ref{tab:crossSectionsData}), with a linear connection from
    the threshold energy, $E_{\mathrm{th}}$, to the peak energy,
    $E_{\mathrm{peak}}$, one obtains:
    
    for $E_{\mathrm{th}} < E \le E_{\mathrm{peak}}$,
    \begin{equation}
	\mathcal{N}_{\gamma}(E_{\mathrm{in}}) = \frac{A}{Z^{2}}
	\frac{2\sigma_{\mathrm{peak}}}{9\sigma_{\mathrm{T}}\ln\Lambda}\,
	\frac{(E_{\mathrm{in}} - E_{\mathrm{th}})^{2}}
	{m_{\mathrm{e}}c^{2}m_{\mathrm{p}}c^{2}}
	\frac{2 E_{\mathrm{in}} + E_{\mathrm{th}}}
	{E_{\mathrm{peak}} - E_{\mathrm{th}}},
	\label{eq:AnalyticFormula1a}
    \end{equation}
    
    and for $E > E_{\mathrm{peak}}$,
    \begin{equation}
	\begin{split}
	\mathcal{N}_{\gamma}&(E_{\mathrm{in}}) =
	\mathcal{N}_{\gamma}(E_{\mathrm{peak}})\\
	&+ \frac{A/Z^{2}}{3\sigma_{\mathrm{T}}\ln\Lambda}\,
	\left[\frac{4E^{2}}{m_{\mathrm{e}}c^{2}m_{\mathrm{p}}c^{2}}
	\left(\frac{aE^{-x}}{2-x} + \frac{b}{2}\right)
	\right]_{E_{\mathrm{peak}}}^{E_{\mathrm{in}}},
	\end{split}
	\label{eq:AnalyticFormula1b}
    \end{equation}
    where we have used the non-relativistic relation between $v$ and
    $E$, which is justified for the energy range under consideration. 
    
    \subsection{High energy limit}
    
    In the other limit, at high energy, the destruction time is so
    small compared to the energy loss time that the particles will be
    destroyed before they have lost any significant amount of energy. 
    We can thus assume that the gamma-ray emission occurs at a
    constant energy, namely the injection energy.  The problem is
    solved straightforwardly in terms of the production and
    destruction timescales (or cross sections):
    \begin{equation}
	\mathcal{N}_{\gamma}(E_{\mathrm{in}}) =
	\frac{\tau_{\mathrm{D}}}{\tau_{\mathrm{prod}}} =
	\frac{\sigma_{\mathrm{prod}}}{\sigma_{\mathrm{D}}}.
	\label{eq:AnalyticFormula2}
    \end{equation}

    Indeed, if one particle is injected at time $t = 0$, the number of
    remaining particles at time $t$ is $N(t) =
    \exp(-t/\tau_{\mathrm{D}})$, and the gamma-ray production rate is
    $\d N_{\gamma}/\d t = N(t)/\tau_{\mathrm{prod}} = 
    \exp(-t/\tau_{\mathrm{D}})/\tau_{\mathrm{prod}}$.  Integrating
    over time, from $t=0$ to $\infty$, gives the above result. 

    \subsection{Accuracy of the analytical formul\ae}

    \begin{figure*}
	\centering
	\includegraphics[width=8cm]{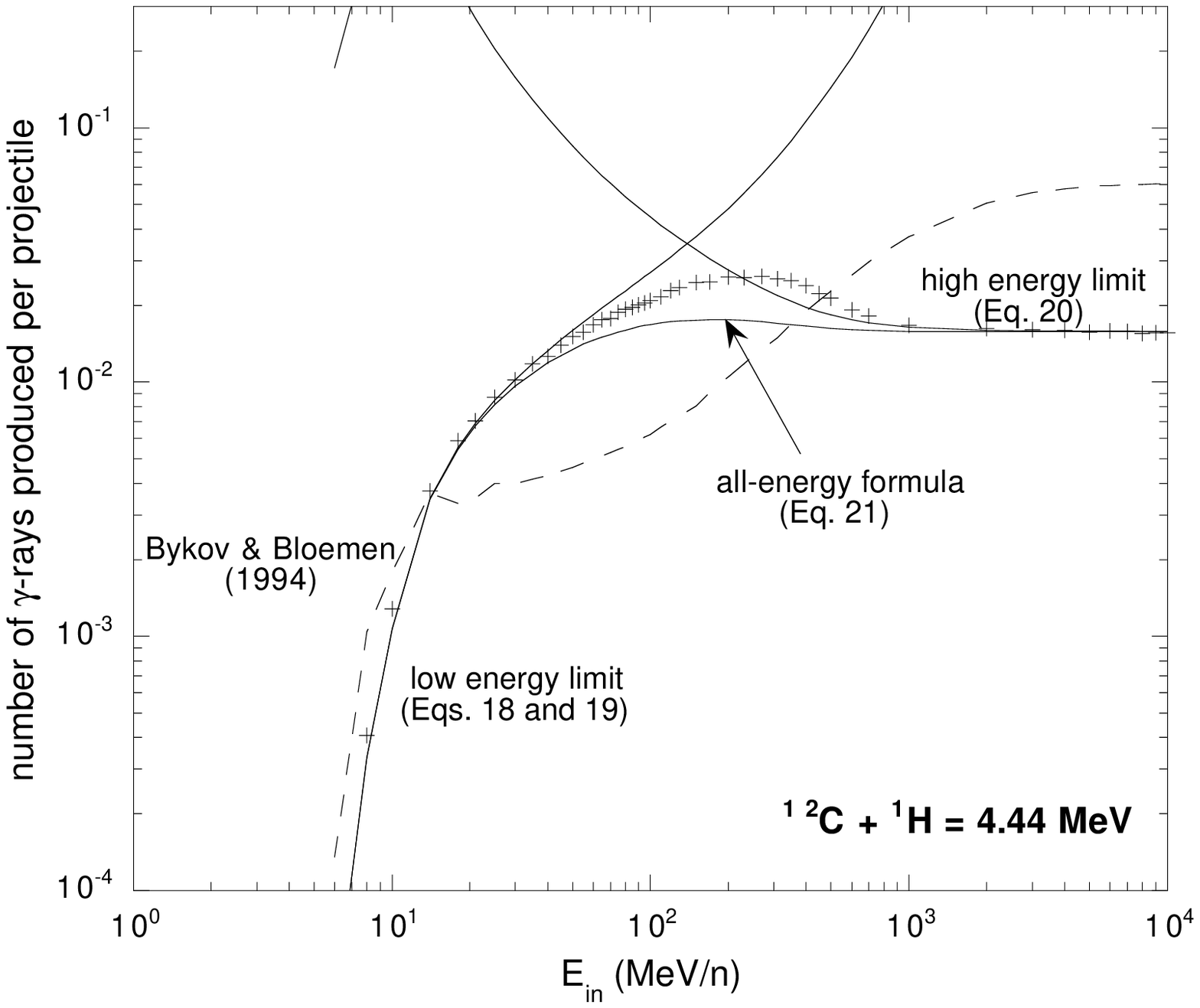}\hspace{0.2cm}
	\includegraphics[width=8cm]{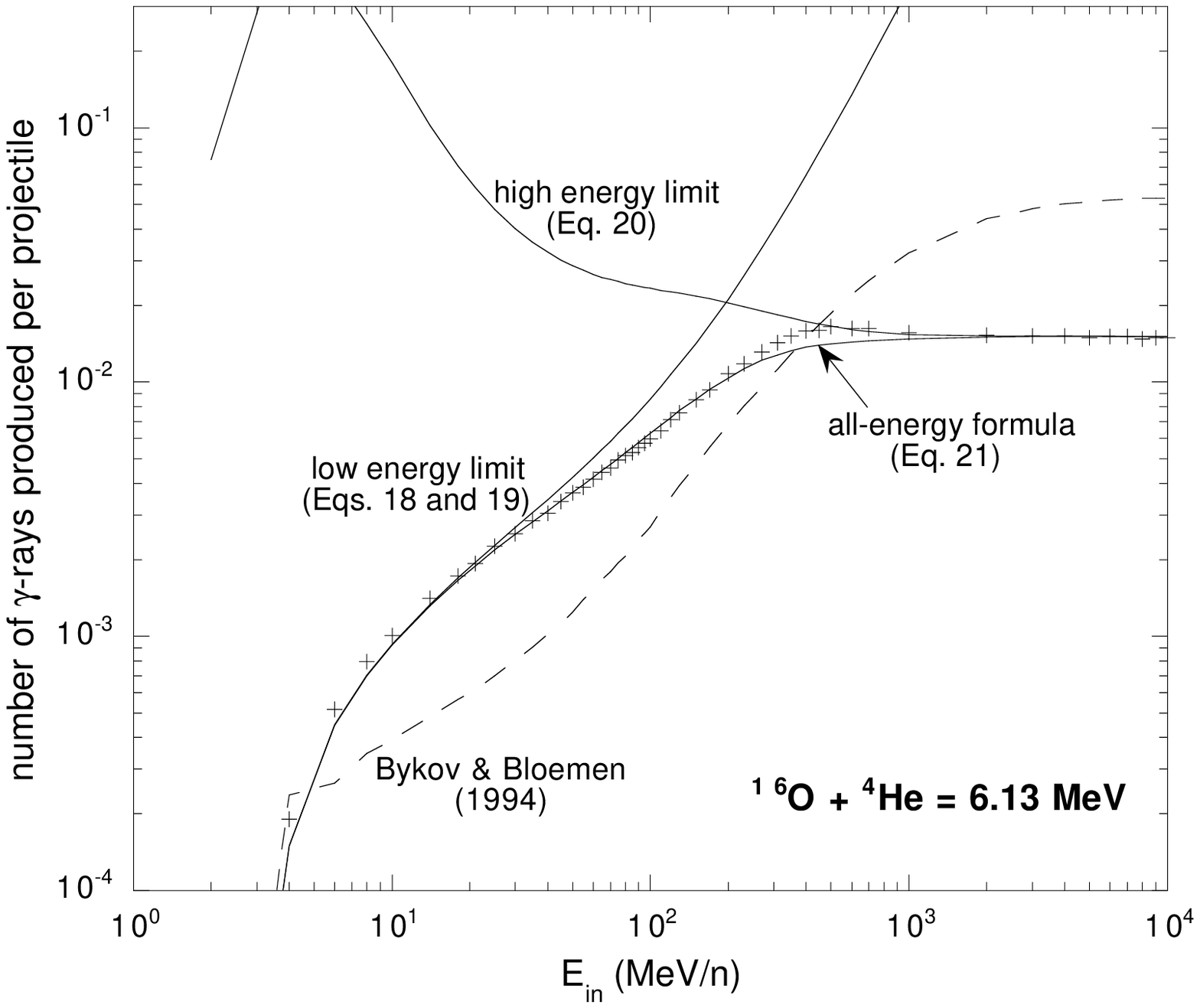}
	\caption{Comparison of the analytical approximations for the
	individual gamma-ray yields with the results of
	Sect.~\ref{sec:Results} (plus signs), for two of the main
	reactions.  The formula of Eq.~(\ref{eq:BykBlo94}) is also
	shown as a dashed line.}
	\label{fig:AnalyticPhotonYields}
    \end{figure*}

    The approached formulae Eqs.~(\ref{eq:AnalyticFormula1a})
    and~(\ref{eq:AnalyticFormula1b}) give an approximation of the
    photon yields at low energy, which may be noted
    $\mathcal{N}_{\gamma,\mathrm{LE}}(E)$, while
    Eq.~(\ref{eq:AnalyticFormula2}) is accurate for the high energy
    limit, $\mathcal{N}_{\gamma,\mathrm{HE}}(E)$.  One can
    then propose an approached formula valid in the whole energy 
    range:
    
    \begin{equation}
	\mathcal{N}_{\gamma}(E) =
	[\mathcal{N}_{\gamma,\mathrm{LE}}(E)^{-1} +
	\mathcal{N}_{\gamma,\mathrm{HE}}(E)^{-1}]^{-1}.
	\label{eq:AnalyticFormula3}
    \end{equation}
    
    In Fig.~\ref{fig:AnalyticPhotonYields}, the various above
    approached formul\ae~are compared with the results of the previous
    section for the reactions $^{12}\mathrm{C} +
    ^{1}\mathrm{H}\rightarrow ^{12}\mathrm{C}^{*}$ and
    $^{16}\mathrm{O} + ^{4}\mathrm{He}\rightarrow
    ^{16}\mathrm{O}^{*}$.  As can be seen, the various approximations
    give reasonably good results in their respective energy range. 
    However, it should be stressed that except for the high energy,
    the accuracy of our formul\ae~depends essentially on the accuracy
    of the cross section modeling.  Since the real cross sections do
    \textit{not} follow the simple analytical expression used in
    Sect.~\ref{sec:LowEnergyLimit}, we cannot expect the analytical
    photon yields calculated in this section to be accurate to more
    than a factor of two or so (as we could observe from the whole
    data set).  This can be very problematic when considering
    gamma-line ratios.  Therefore, we strongly recommend to use the
    results of Sect.~\ref{sec:Results} instead, which are also
    uncertain, but only insofar as the cross sections are not know. 
    They thus provide the best estimates of individual gamma-ray
    yields given the present knowledge on the excitation cross
    sections.

    \section{Summary}
   
    In this paper, we have presented an easier way to calculate
    gamma-ray line emission from energetic particle interactions in
    the ISM. It is based on a simple mathematical transformation whose
    physical interpretation has been given and which allows one to
    work with the source spectrum of the EPs rather than the
    propagated spectrum.  Therefore, one does not need to worry about
    energy-dependent and nucleus-dependent energy losses of the
    particles, nor about their nuclear destruction in-flight, as they
    are taken into account once and for all through the calculation of
    absolute photon yields.  The latter are the number of photons
    produced in each of the nuclear de-excitation lines by a given
    nucleus injected in the ISM at a given initial energy.  These
    photon yields have been given here for various projectiles
    contributing to the main $^{12}$C, $^{14}$N, $^{16}$O, $^{20}$Ne
    and $^{56}$Fe de-excitation lines.  Numerical tables and
    electronic versions of the results are available from the authors
    upon request.  These photon yields are to be used instead of the
    nuclear excitation cross-sections, and might be thought of as
    `effective cross-sections' taking into account the specific
    effects of particle propagation in the ISM. They can be used to
    calculate the $\gamma$-ray line emission induced by EPs with any
    spectrum and any composition in any medium with a metallicity
    lower than a few tens of the solar metallicity.
   
    In addition to simplifying the calculation of gamma-ray line
    emission, the individual EP gamma-ray yields also provide a
    direct, intuitive tool to analyze gamma-ray line data from a
    phenomenological point of view, and construct an EP source
    spectrum and composition which could reproduce the intensity of
    the gamma-ray emission and the various line ratios.  The results
    presented here correspond to a thick target model, which is
    relevant to most astrophysical situations for EPs of energy lower
    than a few hundreds of MeV/n.  However, the same formalism can be
    used to calculate the total EP photon yields in a target with any
    escape length, be it energy-dependent or not.  Once these yields
    have been calculated once, they can be used in any situation with
    the same escape length, for any particle spectrum and any EP and
    target compositions.

    \begin{acknowledgements}
	We wish to thank warmly J\"urgen Kiener in CSNSM Orsay,
	France, for precious comments about energy losses and nuclear
	excitation cross sections.
    \end{acknowledgements}

\end{document}